\documentclass[%
 aip,
 amsmath,amssymb,
 reprint
]{revtex4-1}

\usepackage{graphicx}
\usepackage{dcolumn}
\usepackage{bm}

\usepackage[utf8]{inputenc}
\usepackage[T1]{fontenc}
\usepackage{mathptmx}
\usepackage{etoolbox}

\usepackage[usenames,dvipsnames]{xcolor}
\usepackage[linktoc=page,colorlinks,urlcolor=RedViolet,citecolor=RedViolet,linkcolor=RedViolet]{hyperref}
\usepackage{float}
\usepackage{graphicx}
\usepackage{mathrsfs}
\usepackage{overpic}
\usepackage{wrapfig}
\usepackage{braket}
\usepackage{color}
\usepackage{verbatim}
\usepackage{amssymb,amsmath,mathtools}
\usepackage{upgreek}
\usepackage{bbold}
\usepackage{cancel}
\usepackage{textgreek}
\usepackage[normalem]{ulem}

\makeatletter
\def\@email#1#2{%
 \endgroup
 \patchcmd{\titleblock@produce}
  {\frontmatter@RRAPformat}
  {\frontmatter@RRAPformat{\produce@RRAP{*#1\href{mailto:#2}{#2}}}\frontmatter@RRAPformat}
  {}{}
}%
\makeatother

\newcommand{\UMDphy}{Department of Physics, University of Maryland, College Park, Maryland 20742, USA}
\newcommand{\QTC}{Quantum Technology Center, University of Maryland, College Park, Maryland 20742, USA}
\newcommand{\UMDEECS}{Department of Electrical Engineering and Computer Science,
University of Maryland, College Park, Maryland 20742, USA}
\newcommand{\LincolnLab}{Lincoln Laboratory, Massachusetts Institute of Technology, Lexington, Massachusetts 02421, USA}
\newcommand{\UMDBioE}{Fischell Department of Bioengineering, University of Maryland, College Park, Maryland 20742, USA}

\begin{document}

\title{Quantum Diamond Microscope for Dynamic Imaging of Magnetic Fields}
\date{\today}

\author{Jiashen Tang}
\thanks{These authors contributed equally to this work.}
\affiliation{\UMDphy}
\affiliation{\QTC}

\author{Zechuan Yin}
\thanks{These authors contributed equally to this work.}
\affiliation{\QTC}
\affiliation{\UMDEECS}

\author{Connor A. Hart}
\affiliation{\QTC}
\affiliation{\UMDEECS}

\author{John W. Blanchard}
\affiliation{\QTC}
\affiliation{\UMDEECS}

\author{Jner Tzern Oon}
\affiliation{\UMDphy}
\affiliation{\QTC}

\author{Smriti Bhalerao}
\affiliation{\UMDBioE}

\author{Jennifer M. Schloss}
\affiliation{\LincolnLab}

\author{Matthew J. Turner}
\affiliation{\QTC}
\affiliation{\UMDEECS}

\author{Ronald L. Walsworth}
\email{walsworth@umd.edu}
\affiliation{\UMDphy}
\affiliation{\QTC}
\affiliation{\UMDEECS}

\begin{abstract}
Wide-field imaging of magnetic signals using ensembles of nitrogen-vacancy (NV) centers in diamond has garnered increasing interest due to its combination of micron-scale resolution, millimeter-scale field of view, and compatibility with diverse samples from across the physical and life sciences. Recently, wide-field NV magnetic imaging based on the Ramsey protocol has achieved uniform and enhanced sensitivity compared to conventional measurements. Here, we integrate the Ramsey-based protocol with spin-bath driving to extend the NV spin dephasing time and improve magnetic sensitivity. We also employ a high-speed camera to enable dynamic wide-field magnetic imaging. We benchmark the utility of this quantum diamond microscope (QDM) by imaging magnetic fields produced from a fabricated wire phantom. Over a $270\times270$\,\textmu m$^2$ field of view, a median per-pixel magnetic sensitivity of $4.1(1)$\,nT$/\sqrt{\mathrm{Hz}}$ is realized with a spatial resolution $\lesssim$\,10\,\textmu m and sub-millisecond temporal resolution. Importantly, the spatial magnetic noise floor can be reduced to the picotesla scale by time-averaging and signal modulation, which enables imaging of a magnetic-field pattern with a peak-to-peak amplitude difference of about 300\,pT. Finally, we discuss potential new applications of this dynamic QDM in studying biomineralization and electrically-active cells.
\end{abstract}

\maketitle

\section{Introduction}

\begin{figure*}
    \centering
    \includegraphics[width=\linewidth]{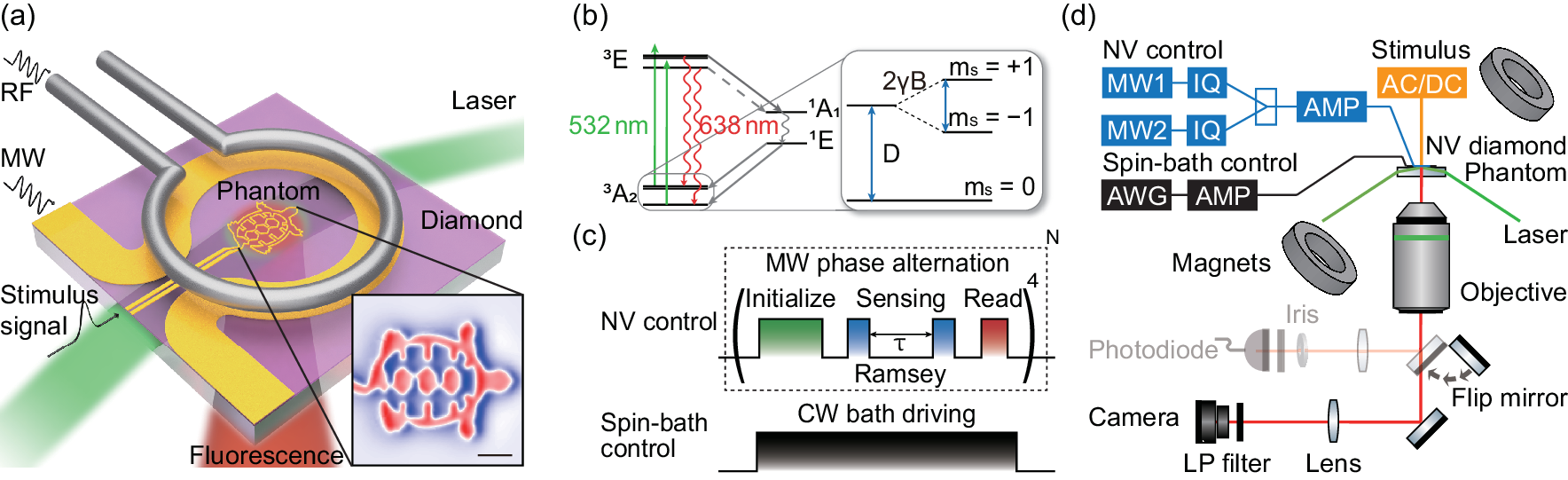}
    \caption{(a) Imaging phantom magnetic fields using a quantum diamond microscope (QDM). 532-nm excitation laser light excites NV centers at the diamond chip surface. A planar, gold omega-loop delivers microwaves (MWs) to the diamond chip for NV spin-state control. Optical components collect NV fluorescence onto either a photodiode or a high-speed camera (not shown). A bias magnetic field (not shown) is applied along one NV axis. An RF signal delivered by a nearby coil drives paramagnetic bath spins in the diamond  to increase the ensemble NV spin dephasing time ($T_2^*$). A driven current in the phantom produces a magnetic-field pattern to be imaged with the QDM. An example image of simulated phantom magnetic signals, projected along the direction of a sensing NV axis with $5$\,\textmu m standoff from the source to the NV surface, is shown in the inset. Scale bar: $50$\,\textmu m.
    (b) NV energy-level diagram. The enlarged view shows the electronic spin triplet ground state with zero-field splitting $D$ and bias magnetic field $B$ aligned to the sensing NV axis. (c) Pulse sequence for magnetic imaging based on integration of Ramsey magnetometry (Top) and spin-bath driving (Bottom). To mitigate NV control error and laser intensity noise, a dual-tone MW is employed and its phase is alternated (see main text).
    (d) Experimental apparatus. 
    A flip mirror can route the NV fluorescence to a photodiode for rapid measurements of ensemble NV spin properties and experimental optimization.
    } 
    \label{fig:1}
\end{figure*}

The nitrogen-vacancy (NV) center in diamond is a solid-state spin defect that emits magnetic-field-dependent fluorescence under optical excitation.
Precision magnetic sensing can be performed at ambient conditions using an ensemble of NVs or a single NV, with wide-ranging applications across the physical and life sciences \cite{levine2019principles, barry2020report, aslam2023quantum}.
In particular, the wide-field magnetic imaging modality known as the quantum diamond microscope (QDM) employs a dense, micron-scale-thick surface layer of NVs on a diamond chip, onto which a sample of interest is placed.
The QDM has been applied for studying both static and dynamic magnetic signals from diverse samples, including ancient rocks and meteorites, 2D condensed matter systems, and electronic circuits \cite{glenn2017micrometer, levine2019principles, ku2020imaging,Turner2020IC}. The QDM is also useful for many life science applications, due to the biocompatibility of the diamond surface.~Examples include characterization of iron mineralization in bacteria \cite{le2013optical}, malarial hemozoin crystals \cite{fescenko2019diamond}, and iron organelles in homing pigeons \cite{de2021quantum}; understanding the microscopic origin of MRI contrast \cite{davis2018mapping}; and tracking the tumbling dynamics of DNA-tethered magnetic particles \cite{Kazi2021wide} among other applications \cite{glenn2015single, mccoey2020quantum,troise2022vitro}. 

A majority of QDMs realized to date utilize a continuous-wave optically detected magnetic resonance (CW-ODMR) sensing protocol. However, competing effects of the optical and microwave (MW) fields employed during CW-ODMR measurements constrain the achievable sensitivity for a fixed number of NVs  per imaging pixel
\cite{barry2020report}.
This limitation results in averaging intervals of hours for nanotesla  magnetic fields \cite{le2013optical, de2021quantum, davis2018mapping}, restricting the imaging of weaker magnetic sources, as well as sample throughput.~The demand for sensitivity is especially critical when both spatial and time resolution are required for imaging dynamic magnetic fields, such as from action potential (AP) currents in electrically-active cells.~To date, dynamic NV-based measurements have detected magnetic fields from neuron \cite{barry2016optical}, cardiac \cite{arai2022millimetre}, and muscle \cite{troise2022vitro} AP currents, but only by spatially integrating the ensemble NV fluorescence (over $>$\,100\,\textmu m length scales) onto a single photodiode and signal averaging over multiple AP measurements.~In particular, the temporal resolution in demonstrated QDM dynamic biomagnetic imaging experiments \cite{davis2018mapping,Kazi2021wide} are inadequate compared to the sub-millisecond timescales required for resolving individual AP currents.

QDM per-pixel magnetic sensitivity and temporal resolution can be improved by implementing a pulsed sensing protocol with acquisition using a high-frame-rate camera \cite{Hart2021-4Ramsey,turner2020huphd,schloss2019optimizing}. 
Pulsed protocols using Ramsey magnetometry, in particular, separate intervals of NV optical preparation and readout, MW control, and sensing, affording optimization of each NV-interrogation stage for improved sensitivity compared to CW-ODMR. Recent advances in Ramsey-based magnetic imaging protocols have demonstrated robustness to errors from heterogeneous MW control fields, diamond strain, and temperature \cite{Hart2021-4Ramsey}, enabling spatially uniform and order-of-magnitude improved sensitivity compared to CW-ODMR-based measurements \cite{Turner2020IC}.

In this work, we benchmark the utility of a
Ramsey-based QDM using a high-speed, lock-in camera by imaging static and dynamic magnetic fields from a fabricated wire phantom (Fig.~\ref{fig:1}(a)). The phantom's geometry is designed to generate nontrivial, micron-scale spatial patterns of magnetic fields. 
We additionally integrate spin-bath driving \cite{bauch2018ultralong}
to  
improve the magnetic sensitivity by extending the NV spin dephasing time, expanding the utility of this technique from previous confocal-volume demonstrations to wide-field imaging. After describing the experimental setup in Section~\ref{sec:methods}, we first characterize the performance of the magnetic imaging system, including analysis of sensitivity in Section~\ref{sec:sensitivity} and spatial magnetic noise floor in Section~\ref{sec:sigmaspatial}. Next, we present static magnetic-field imaging experiments with micron-scale spatial resolution in Section~\ref{sec:dc}. 
Importantly, we demonstrate the capability to resolve a picotesla-scale magnetic-field pattern by time-averaging and signal modulation; and show the usefulness of a denoising technique \cite{buades2011non} in improving the image signal-to-noise ratio (SNR).
In Section~\ref{sec:dynamic}, we image dynamic magnetic fields by applying to the phantom a broadband, synthetic human cardiac signal. We show that a time series of QDM magnetic images can capture temporal variations with sub-millisecond time resolution. 
Given these demonstrations, we discuss potential applications in imaging static and dynamic biomagnetic signals in Section~\ref{sec:discussion}, including magnetic characterization of iron-loaded compartments in engineered eukaryotic cells \cite{sigmund2018bacterial}, high-throughput screening of biogenic magnetite across candidate tissues for vertebrate magnetoreceptors \cite{diebel2000magnetite, eder2012magnetic, edelman2015no, de2021quantum}, and monitoring of
currents in cardiomyocytes. 
Finally, we provide an outlook 
towards imaging
weaker and more transient biomagnetic fields such as signals produced by neuronal currents.

\section{EXPERIMENTAL METHODS}
\label{sec:methods}

The NV center
is 
a $C_{3v}$-symmetric defect center 
formed by substitution of a nitrogen atom adjacent to a vacancy in the
diamond lattice. 
Its electronic spin $S=1$ ground state (Fig.~\ref{fig:1}(b)) has a zero-field splitting $D=2\pi\times2.87$\,GHz at room temperature, separating the $\ket{m_s=0}$ and $\ket{m_s=\pm1}$ sublevels. Application of a bias magnetic field further splits the degenerate $\ket{m_s=+1}$ and $\ket{m_s=-1}$ states through the Zeeman effect.
For the present QDM experiments, a nominal 4.3\,mT bias magnetic field is aligned to one of the four NV ensemble 
axes
in diamond for sensing the projection of signal magnetic fields along that particular sensing NV axis (defined as the $z$ axis).
Under this condition,
transverse crystal stress and electric fields
can be neglected in the ground-state sensing NV Hamiltonian \cite{Hart2021-4Ramsey, Doherty2013,Kehayias2019stress, Udvarhelyi2018spinstrain}, resulting in the approximate form 
\begin{equation}
\label{eq:NVHamiltonian}
\begin{aligned}
\hat{H}/\hbar \approx  [D+M_z] \hat{S}_z^2+\gamma B_z\hat{S_z},
\end{aligned}
\end{equation}
where $\hat{S_z}$ is the dimensionless spin-1 operator, $\gamma = 2\pi\times 28.024$\,GHz/T \cite{Acosta2010} is the NV electron gyromagnetic ratio, and $B_z$ and $M_z$ are longitudinal components of the bias magnetic field and crystal stress, respectively. (More generally, vector magnetic sensing is possible by measuring the components of the signal field along all NV axes \cite{barry2020report}.)

The sequence employed for magnetic imaging is shown in Figure~\ref{fig:1}(c), and the experimental apparatus is shown schematically in Figure~\ref{fig:1}(d). The diamond employed in these demonstration experiments has a 10\,\textmu m NV ensemble layer 
($[\text{NV}]$~$=$~2.4\,ppm, $^{15}$N enriched) on a $2\times2\times0.5$\,mm$^3$ substrate ($>$\,99.995\% $^{12}$C).
During the Ramsey sequence, 
a pulse of 532\,nm laser irradiation first initializes the NV electronic spin state to $\ket{m_s=0}$ via optical pumping.
Then, a dual-tone MW pulse
prepares the NV spin state as a superposition between $\ket{m_s=+1}$ and $\ket{m_s=-1}$ for magnetic sensing. In the subsequent Ramsey free evolution interval $\tau$, the presence of an additional signal magnetic field $B_{sig}$ along the sensing NV axis causes accumulation of a relative phase $\phi_{B}=2\gamma B_{sig}\tau$ between $\ket{m_s=+1}$ and $\ket{m_s=-1}$ in the rotating frame \cite{mamin2014multipulse}. 
The optimal choice \cite{barry2020report} of $\tau$ is limited by the NV spin dephasing time $T_2^*$. 
To extend $T_2^*$ while retaining sensitivity to static and broadband magnetic signals, RF control fields resonant with the paramagnetic spin-bath transitions in the diamond are applied during the free evolution interval. This spin-bath-driving technique decouples unwanted dipolar interactions between NV sensor spins and paramagnetic bath spins (Section~\ref{supp:DEER} in supplementary material). 
At the end of the Ramsey free evolution interval, a second dual-tone MW pulse maps the accumulated phase information $\phi_{B}$ into a population difference between $\ket{m_s=0}$ and $\ket{m_s=\pm1}$. This population difference, proportional to the magnetic signal $B_{sig}$, is subsequently read out using the spin-state-dependent NV fluorescence via a lock-in camera (Heliotis heliCam C3) that is capable of external frame rate of up to 3.8\,kHz.
The camera has a tunable internal exposure frequency of up to 1\,MHz, which is synchronized with MW-phase-alternated Ramsey measurements. This particular variation of the Ramsey sequence is designed to mitigate laser intensity noise and NV spin-control errors in the accumulated signal contained in each external frame (Section~\ref{supp:DQ4R} in supplementary material).
A $270\times270$\,\textmu m$^2$ region of the NV layer is selected as the field of view, where each pixel corresponds to a lateral area of about $1.9\times1.9$\,\textmu m$^2$. In the following experiments, up to $3\times3$ pixel binning is applied such that the magnetic spatial resolution is expected to be limited by the 10-\textmu m-thick NV layer.  
Additional details regarding the experimental setup are provided in Section~\ref{supp:exp} of supplementary material.

\section{RESULTS}
\subsection{NV spin dephasing time and magnetic sensitivity}
\label{sec:sensitivity}
\begin{figure}[t]
    \centering
    \includegraphics[width=\linewidth]{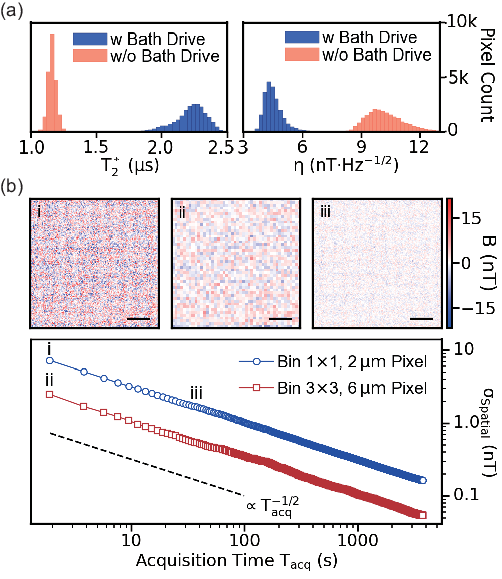}
    \caption{
    Measured magnetic sensing performance of the Ramsey-based QDM. (a) Median per-pixel $T_2^*$ is 2.2(1)\,\textmu s and 1.2(1)\,\textmu s with and without spin-bath driving. Median per-pixel $\eta$ is 4.1(1)\,nT$/\sqrt{\mathrm{Hz}}$ and 9.4(1)\,nT$/\sqrt{\mathrm{Hz}}$ with and without spin-bath driving.
    (b) Spatial magnetic noise floor. Measurement data are obtained using the differential measurement protocol (see main text). Top: Images of magnetic noise with different acquisition times and pixel binning (see the bottom plot). Here, the acquisition time accounts for the duration of time spent acquiring measurements, but excludes the implementation-dependent time required to transfer and store data from the camera to a host computer. The magnetic noise is distributed randomly without significant spatial correlation. Field of view: $270\times270$\,\textmu m$^2$. Scale bar: 50\,\textmu m. Bottom: Spatial magnetic noise floor $\sigma_{spatial}$ as a function of acquisition time $T_{acq}$ for no pixel binning and $3\times3$ pixel binning. $\sigma_{spatial}$ is computed as the standard deviation of magnetic-field measurements using the entire image after any pixel binning. A dashed black line depicts power law scaling behavior $\propto T_{acq}^{-1/2}$ as a guide to the eye.
    }
    \label{fig:2}
\end{figure}

We first characterize the performance of the Ramsey-based QDM with and without spin-bath driving. The NV spin dephasing time $T_2^*$ and magnetic-field sensitivity $\eta$ are studied on a pixel-by-pixel basis over the field of view.
For these measurements, a dual-tone MW implementation of the Ramsey sequence leverages a strain- and temperature-insensitive coherence (Section~\ref{supp:DQ4R} in supplementary material) such that the NV ensemble $T_2^*$ is limited by dipolar interactions with bath spins. Under this condition, spin-bath driving improves the NV spin dephasing time and magnetic sensitivity as shown in Figure~\ref{fig:2}(a).
The median per-pixel $T_2^*$ is extended by about 1.8$\times$ to 2.2(1)\,\textmu s, approaching the estimated NV-NV interaction limit of 2.5\,\textmu s for this sample (Section~\ref{supp:sensorcharacterization} in supplementary material). The median per-pixel $\eta$ is enhanced by about
2.4$\times$ to 4.1(1)\,nT$/\sqrt{\mathrm{Hz}}$ in the absence of pixel binning; and is comparable to  
the combined quantization and photon-shot-noise-limited sensitivity estimate of 3.2\,nT$/\sqrt{\mathrm{Hz}}$ (in roughly equal contribution, Section~\ref{supp:sensorcharacterization} in supplementary material).
For photon shot-noise-limited Ramsey magnetometry, the magnetic sensitivity can be written as \cite{barry2020report} (also see Equation~\ref{suppeq:RamseyPhotonShotSensitivity} in supplementary material) 
$\eta_{shot}\propto \sqrt{t_{D}+\tau}/\tau Ce^{(-\tau/T_2^*)^p}$
, where $C$ is the NV spin-state readout contrast, $\tau$ is the Ramsey free evolution interval, $p$ is a parameter used to describe the Ramsey envelope decay shape, and $\tau_D$ is the overhead time for NV initialization and readout.
The present QDM operates in a regime where the sensitivity nominally improves linearly with increased dephasing time \cite{barry2020report, bauch2018ultralong}, as $\tau_D=7.04$\,\textmu s is longer than the sensing interval $\tau\approx T_2^*$, yielding 
$\eta_{shot}\propto \sqrt{t_{D}}/\tau C$
.
The additional superlinearity of the measured $\eta$ improvement is attributed to an observed increase in spin-state readout contrast $C$ of about $1.2\times$ when using spin-bath driving, and the discrete choices of $\tau$ due to Ramsey fringe beating introduced by the hyperfine splitting of the NV spin resonances \cite{Hart2021-4Ramsey}. 

The measured per-pixel magnetic sensitivity describes the expected magnetic noise after 1\,s of signal acquisition. Here, the signal acquisition time $T_{acq}$ accounts for the duration of time allocated to acquiring measurements, but excludes the implementation-dependent duration required to transfer and store the data from the 500-frame camera buffer to a host computer (Section~\ref{supp:exp} in supplementary material). This technical overhead is included below when reporting a "wall-clock" time $T_{wall}$. The frame rates, wall-clock times, and Allan deviations associated with magnetic sensitivity results in Figure~\ref{fig:2}(a) are included in Section~\ref{supp:sensorcharacterization} of supplementary material. 

\subsection{Spatial magnetic noise floor}
\label{sec:sigmaspatial}
In magnetic imaging, the ability to resolve a signal of interest depends upon the per-pixel magnetic sensitivity and spatial magnetic noise floor \cite{glenn2017micrometer}. Similar to Ref.~\onlinecite{glenn2017micrometer}, we quantify the spatial magnetic noise floor $\sigma_\text{spatial}$ after a certain acquisition time, by calculating the standard deviation of magnetic-field measurements across the time-averaged image 
without any external signal sources present. Uncorrelated spatial noise can be time-averaged. However, correlated spatial noise across an image, particularly when similar to the length scale of the signal of interest, inhibits the utility of temporal averaging and limits the minimum resolvable magnetic field. Specifically, vibration and temperature changes can induce spatial and temporal variations in the bias magnetic field and optical illumination intensity, thereby introducing correlated noise into QDM magnetic images. 

For QDM magnetic images of an external signal source (such as the phantom used in the present demonstration experiments), we mitigate the impact of spatially correlated noise varying slowly on the timescale between acquisitions of frame sets (i.e., the time required to transfer the collected 500 frames to the host computer) by employing a differential measurement protocol. Currents in the phantom are modulated after one acquisition of the 
500-frame set by either on-off gating or reversing the polarity (Section~\ref{supp:diffmeas} in supplementary material). This protocol enables long-term averaging and is applicable to studies of signals that can be externally controlled, e.g., from electronic circuits and electrically-active biological cells.
When studying signal sources that cannot be varied straightforwardly, alternative protocols such as modulating the polarity of bias magnetic field, more advanced MW pulse sequence schemes, or 
post-processing filtering can be employed to mitigate the spatially correlated noise \cite{fu2023pinpointing, Marshall2022precision, barry2016optical}.

As shown in Figure~\ref{fig:2}(b), the spatial distribution of noise in QDM images, at different acquisition times $T_{acq}$ and with no current flowing in the phantom, have insignificant correlation when using the differential measurement protocol. The spatial magnetic noise floor, $\sigma_{spatial}$, can be averaged down as reflected in the observed $\sigma_{spatial}\propto T_{acq}^{-1/2}$ scaling persistent up to about $T_{acq}=4000$\,s ($T_{wall}\approx8$\,h). An inverse proportionality $\sigma_{spatial}\propto1/n_{bin}$ is observed as $n_{bin}\times n_{bin}$ pixels are averaged, allowing 
spatial resolution to be traded for improved spatial magnetic noise floor and reduced acquisition time. Additional spatial magnetic noise floor measurements, for different pixel binning and without employing the differential
measurement protocol, are included in Section~\ref{supp:diffmeas} and \ref{supp:notmodulated} of supplementary material, respectively.

\subsection{Static magnetic-field imaging}
\label{sec:dc}
\begin{figure}[t]
    \centering
    \includegraphics[width=\linewidth]{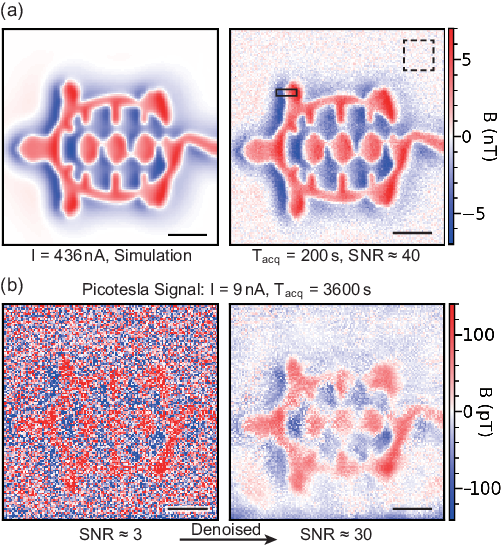}
    \caption{
     (a) Simulated (Left) and measured (Right) phantom static magnetic-field patterns using the Ramsey-based QDM and a  differential measurement protocol. The measurement is conducted with a 25\,mV voltage (436\,nA current) applied to the phantom and time-averaged for about 200\,s. To compute SNR, the peak-to-peak amplitude difference of the magnetic-field pattern (solid box) is divided by the spatial magnetic noise obtained from a source-free region (dashed box). The simulated phantom magnetic-field pattern is obtained using the applied current and geometry of the phantom. To mimic the parameters of the imaging system, the simulated magnetic fields are further averaged over a depth of $10$\,\textmu m (the NV-layer thickness) and binned with a pixel area of $1.9\times1.9$\,\textmu m$^2$. NV-phantom standoff distance is the only free parameter, and is tuned to a value of 5\,\textmu m to best match the measured magnetic-field pattern. 
     (b) Demonstration of a picotesla-scale QDM magnetic image. Left: After about $T_{acq}=3600$\,s ($T_{wall}\approx7$\,h) time-averaging, a magnetic-field pattern with a peak-to-peak amplitude difference of about 300\,pT is resolved with $\mathrm{SNR}\approx3$. 
     Right: Applying a non-local mean denoising technique enhances image $\mathrm{SNR}$ by about 10$\times$. 
     All images are not binned and scale bars are 50\,\textmu m for (a, b).}
    \label{fig:3}
\end{figure}

We characterize the capability of the Ramsey-based QDM to image static magnetic fields by applying steady currents to the fabricated wire 
phantom. The polarity of applied signals is reversed between acquisitions of frame sets to allow differential measurements (Section~\ref{supp:diffmeas} in supplementary material).
As shown in Figure~\ref{fig:3}(a), an imaged magnetic-field pattern with a peak-to-peak amplitude difference of about 14\,nT is obtained by supplying a fixed voltage of $25$\,mV (436\,nA current) to the phantom and averaging for about $T_{acq}=200$\,s ($T_{wall}\approx21$\,min). 
We simulate the expected phantom magnetic-field pattern by using finite element software \cite{comsol} to  
calculate
the current density distributions and the accompanying magnetic signals given the phantom geometry and experimental input current $I$. 
The standoff distance $d_{so}$ between the phantom and NV surface is the only free parameter in the simulation and is chosen to best match the measured magnetic-field pattern \cite{kehayias2022measurement}, yielding $d_{so}\approx5$\,\textmu m. Additional simulated phantom magnetic-field patterns projected along different NV crystal axes are shown in supplementary Section~\ref{supp:fouraxes}.

We also demonstrate QDM magnetic imaging of a picotesla-scale signal 
by applying a $500$\,\textmu V voltage ($9$\,nA current) to the phantom. After averaging for about $T_{acq}=3600$\,s ($T_{wall}\approx7$\,h), a phantom magnetic-field pattern with a peak-to-peak amplitude difference of about 300\,pT is resolved with $\mathrm{SNR}\approx3$ as shown in Figure~\ref{fig:3}(b), left panel. 
In addition, as the magnetic noise has insignificant pixel-to-pixel correlation (Fig.~\ref{fig:2}(b)), we can apply image denoising techniques to enhance SNR. A non-local mean denoising method \cite{buades2005non, buades2005review,  buades2011non} (Section~\ref{supp:denoise} in supplementary material) implemented in open-source software \cite{opencv_library} enables an order-of-magnitude improvement of SNR, as seen in Figure~\ref{fig:3}(b), right panel. 
 
\subsection{Dynamic magnetic-field imaging}
\label{sec:dynamic}
\begin{figure}[t]
    \centering
    \includegraphics[width=\linewidth]{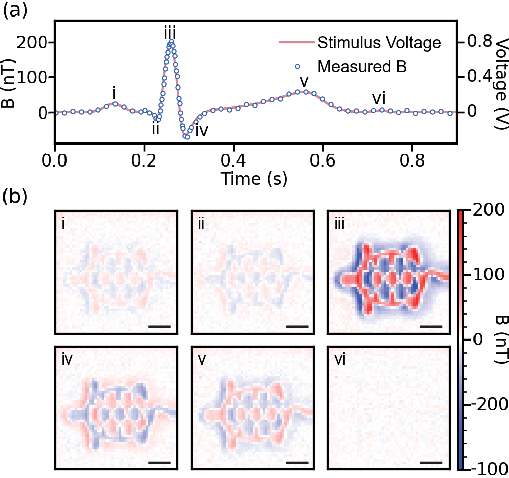}
    \caption{Time-resolved imaging of dynamic magnetic fields using the Ramsey-based QDM and a differential measurement protocol. A voltage trace mimicking a synthetic, broadband human cardiac signal is applied to the phantom. Sets of QDM magnetic images are collected with temporal resolution of about 2\,ms and binned with $3\times3$\,pixels. The experiments are repeated 400 times to allow signal averaging. (a) Temporal variation of the peak magnetic field from the signal-averaged images is overlaid with the applied stimulus. A voltage-to-magnetic-field scaling factor of 274.26\,nT/V is applied to the entire voltage trace (see main text). A subset of peak magnetic-field amplitudes from the acquired images are displayed, indicating good agreement between the applied voltages and magnetic measurements.
    (b) Selected magnetic-field images at time points labeled in (a). Each image has been signal averaged. The spatial magnetic noise at time (vi) is about 3\,nT.
    } 
    \label{fig:4}
\end{figure}

Access to individual frames from the QDM's lock-in camera 
permits imaging of dynamic magnetic signals. As an example, we apply a 1\,s-long 
broadband voltage trace, mimicking a human cardiac signal, to the phantom; and image the associated temporal variations of the magnetic-field patterns. 
The camera external frame rate $F_s$ is set to about $0.5$\,kHz to maximize time resolution while balancing the 500-frame camera buffer limitation on a continuous acquisition (Section~\ref{supp:exp} in supplementary material). The signals applied to the phantom are gated between on and off across successive acquisitions of frame sets (Section~\ref{supp:diffmeas} in supplementary material).
The experiment is repeated $400$ times.

Figure~\ref{fig:4} displays the imaged dynamic magnetic fields with $3\times3$ pixel binning (spatial resolution $\lesssim$\,10\,\textmu m). The applied voltage trace is shown in Figure~\ref{fig:4}(a) and overlaid with the measured peak magnetic-field amplitude.
To demonstrate the capability of high-fidelity waveform-reconstruction, the entire voltage trace is multiplied with a voltage-to-magnetic-field scaling factor obtained from the applied voltage and measured peak magnetic-field amplitude data in the static imaging experiment (Fig.~\ref{fig:3}(a)).
The good agreement shown in Figure~\ref{fig:4}(a) motivates possible applications such as comparison with patch clamp electrophysiology recording. 
A series of magnetic images at time points corresponding to the pseudo-cardiac signal local extrema are shown in Figure~\ref{fig:4}(b). 
The phantom magnetic-field pattern becomes indistinguishable from the background at (vi), which is used to obtain the spatial magnetic noise floor $\sigma_{spatial}\approx3$\,nT.

Magnetic imaging with sub-millisecond time resolution is also feasible with the present QDM, as the external frame rate $F_s$ can be increased to 3.8\,kHz. As an example, single-frequency oscillating voltages ($<$ Nyquist frequency, $F_s/2$) are applied to the phantom, with measured results shown in Section~\ref{supp:subms} of supplementary material.

\section{DISCUSSION}
\label{sec:discussion}
We demonstrate a quantum diamond microscope (QDM) for sensitive, high-speed wide-field imaging of static and broadband magnetic signals, based on Ramsey magnetometry and spin-bath-driving techniques. The dual-tone MW implementation of the Ramsey sequence mitigates sensitivity degradation due to heterogeneous MW control fields and diamond strain; while spin-bath driving further extends the NV spin dephasing time and improves magnetic sensitivity by decoupling NV interactions with bath spins. Over a $270\times270$\,\textmu m$^2$ field of view, a median per-pixel magnetic sensitivity of 4.1(1)\,nT$/\sqrt{\mathrm{Hz}}$ is realized; and the spatial magnetic noise floor can be averaged down to the picotesla scale using a differential measurement protocol. In addition, the QDM's lock-in camera permits time-resolved magnetic imaging with sub-millisecond temporal resolution.

The Ramsey-based QDM presented in this work is a key step towards applications in the physical and life sciences. For example, magnetic-particle-based studies using previous QDMs 
have probed nanotesla signal amplitudes
\cite{glenn2015single,davis2018mapping, Kazi2021wide}.
The sensitivity achieved with the present QDM should allow studies of picotesla-scale signal amplitudes. As an example of detecting weak biomagnetic signals, iron-loaded bacterial encapsulin compartments expressed in genetically engineered mammalian cells are of interest \cite{sigmund2018bacterial}. Magnetization in such 30\,nm-diameter shell structures has been demonstrated to produce beneficial $T_2$ contrast in MRI. Cellular engineering to enhance iron loading, and, in particular, to screen for superparamagnetism, can facilitate optimization of such MRI contrast agents. 
Assuming volumetric magnetization similar to commercial superparamagnetic iron-oxide nanoparticles \cite{davis2018mapping}, an engineered 30\,nm-diameter compartment is expected to produce a magnetic field $\sim$\,600\,pT at a 10\,\textmu m standoff, which can be resolved by the Ramsey-based QDM within a reasonable averaging time ($\sim$\,100\,s of signal acquisition).

The Ramsey-based QDM could also function as a high-throughput biomagnetism tool, such as searches for biogenic magnetite particles related to vertebrate magnetoreception \cite{winklhofer2010quantitative} or population-based measurements.
The localization of cells containing magnetite has been challenging, as organism-scale volumes of candidate tissues need to be screened. Successful identification of magnetite in salmon \cite{diebel2000magnetite, walker1997structure} using magnetic force microscopy suggests an order-of-magnitude estimate for the dipole moment of $10^{-16}$\,A\,m$^2$.
Assuming magnetite in other model animals has similar dipole moments, the magnetic field magnitude immediately outside a $10$\,\textmu m cell is expected to be tens of nanotesla, which can be imaged by the present QDM with $\mathrm{SNR}\approx10$ in a few minutes of signal averaging (see Fig.~\ref{fig:3}(a)). We estimate a tissue area magnetite screening rate of a few mm$^2$ per hour using the present QDM, assuming a 10\,\textmu m-thick tissue. In addition, the QDM's ability to measure  spatial distributions of vector magnetic fields enables quantitative estimations of dipole moments \cite{davis2018mapping, le2013optical}. We thus expect the QDM may become a promising tool for the rapid screening of vertebrate magnetoreceptor cells.

Beyond imaging of static magnetic fields, the Ramsey-based QDM's capability to measure temporal dynamics can benefit studies of electrically-active cells. For example, spatially resolved measurements of electrical currents in cardiac tissue --- via imaging of induced dynamic magnetic fields --- could inform biophysics modeling at the cellular level \cite{jaeger2022deriving, rohr2004role, dhein2021remodeling} or aid in pharmaceutical studies \cite{barral2016utility}. 
Previous 
measurements of the heart have recorded nanotesla-scale magnetic fields at $>$\,100\,\textmu m standoff distances \cite{Holzer2004-magneticimagecardiac, arai2022millimetre}.  
As the NV sensing layer can be brought to few micrometer standoff distances from live bio-samples \cite{barry2016optical}, and externally controlled currents ($\gtrsim$\,1\,\textmu A) can be applied to cardiac tissue \cite{rossi2017effect}, we expect the Ramsey-based QDM may find applications in mapping the microscopic electrical properties of such tissue \cite{sepulveda1989current,rossi2017incorporating}.

The sub-millisecond time resolution of the present QDM (Section~\ref{supp:subms} in supplementary material) is sufficient for mapping the dynamic magnetic fields produced by neuronal action potential (AP) currents \cite{Hochbaum2014-rhodopsins}.
However, resolving the expected $\lesssim$\,1\,nT AP magnetic signal \cite{sundaram2016direct} with a $\sim$\,0.5\,ms temporal resolution requires $\gtrsim$\,4\,k averages ($\gtrsim$\,2\,day of wall-clock time) to reduce the spatial magnetic noise floor $\sigma_{spatial}$ to $\lesssim$\,0.8\,nT using $5\times5$ pixel binning ($\sim$\,$10\times10\times10$\,\textmu m$^3$ sensing volume) with the present sensitivity and technical overhead time. While repeated AP excitation of cultured neurons $in\ vitro$ is feasible (continuous stimulation at 4\,Hz over days has been reported \cite{ihle2022experimental}), the long experimental time poses a significant challenge. 
For the QDM to become an attractive tool for this application, at least an order-of-magnitude improvement in volume-normalized (i.e., per-pixel) sensitivity is likely required. In the present system, increasing the optical illumination intensity for improved NV fluorescence signal results in degraded contrast due to NV charge-state conversion \cite{edmonds2021characterisation}. The optimal laser illumination intensity found in the present work $\sim$\,0.014\,mW/\textmu m$^2$ is far from the NV saturation intensity \cite{wee2007two} ($1-3$\,mW/\textmu m$^2$), highlighting the importance of further diamond engineering to mitigate charge-state issues. The ensemble NV spin dephasing time after spin-bath driving $T_2^*=2.2$(1)\,\textmu s is limited by NV-NV dipolar interactions. Homonuclear-decoupling techniques \cite{balasubramanian2019dc} that mitigate NV-NV interactions while maintaining sensitivity to static and broadband magnetic fields are promising for further NV $T_2^*$ extension.
Beyond the direct sensitivity improvements due to prolonged NV spin dephasing time, improving the sensing duty cycle in the pulse sequence (i.e., increasing the fraction of time spent in the Ramsey free evolution interval) may
warrant the use of alternative readout schemes \cite{barry2020report}; and may require further advances in camera and other hardware capabilities. In particular, reducing the overhead time associated with data transfer from the camera buffer to the host computer will be a key challenge for future QDM optimization.

While we focus in this work on imaging the projected component of magnetic signals along a single sensing NV orientation, vector magnetic-field imaging can be also realized by sequentially interrogating the four different NV axes \cite{levine2019principles}. In addition, pulsed protocols for NV AC magnetometry \cite{devience2015nanoscale} may enable wide-field imaging of thermally-polarized NMR signals with micron-scale resolution. Quantum logic enhanced (QLE) techniques for NV ensembles have recently demonstrated $>$\,10$\times$ $\mathrm{SNR}$ enhancement for AC signals \cite{arunkumar2023quantum}. QLE sensing may also be incorporated in future QDM realizations.

\section*{SUPPLEMENTARY MATERIAL}
Supplementary material is included for additional details of
experimental methods.

\begin{acknowledgments}
We thank David R. Glenn and Pauli Kehayias for helpful comments and discussion in preparing the manuscript. We acknowledge the support of the Maryland NanoCenter and its FabLab. This work is supported by, or in part by, the U.S. Army Research Laboratory under Grant No.~W911NF1510548 and Contract No. W911NF1920181; the U.S. Army Research Office under Grant No. W911NF2120110; the DARPA DRINQS program under Grant No. D18AC00033; the National Science Foundation (NSF) Center for Integration of Modern Optoelectronic Materials on Demand (IMOD) via grant DMR-2019444; the U.S. Air Force Office of Scientific Research under Grant No. FA9550-22-1-0312; the Gordon \& Betty Moore Foundation; and the University of Maryland Quantum Technology Center.
\end{acknowledgments}

\section*{AUTHOR DECLARATIONS}
\subsection*{Conflict of Interest}
Ronald L. Walsworth is a founder of and advisor to companies that are developing and commercializing NV-diamond technology.  These relationships are disclosed to and managed by the University of Maryland Conflict of Interest Office.  
\subsection*{Author Contributions}
{\bf Jiashen Tang:} Investigation (Equal); Data curation (lead); Formal analysis (equal); Software (equal); Writing --- original draft (lead); Writing --- review \& editing (equal).
{\bf Zechuan Yin:} Investigation (Equal); Visualization (lead); Formal analysis (equal); Validation (equal); Software (equal); Writing --- review \& editing (equal).
{\bf Connor A. Hart:} Conceptualization (equal); Resources (equal); Validation (equal); Writing --- review \& editing (equal).
{\bf John W. Blanchard:} Resources (equal); Validation (equal); Writing --- review \& editing (equal).
{\bf Jner Tzern Oon:} Software (equal); Validation (equal); Writing --- review \& editing (equal).
{\bf Smriti Bhalerao:} Validation (equal); Writing --- review \& editing (equal).
{\bf Jennifer M. Schloss:} Conceptualization (equal); Writing --- review \& editing (equal).
{\bf Matthew J. Turner:} Conceptualization (equal); Writing --- review \& editing (equal).
{\bf Ronald L. Walsworth:} Funding acquisition (lead); Resources (equal); Supervision (lead); Writing --– review \& editing (equal).

\section*{DATA AVAILABILITY}
The data that support the findings of this study are available from the corresponding author upon reasonable request.

\bibliography{references.bib}

\begin{thebibliography}{58}%
\makeatletter
\providecommand \@ifxundefined [1]{%
 \@ifx{#1\undefined}
}%
\providecommand \@ifnum [1]{%
 \ifnum #1\expandafter \@firstoftwo
 \else \expandafter \@secondoftwo
 \fi
}%
\providecommand \@ifx [1]{%
 \ifx #1\expandafter \@firstoftwo
 \else \expandafter \@secondoftwo
 \fi
}%
\providecommand \natexlab [1]{#1}%
\providecommand \enquote  [1]{``#1''}%
\providecommand \bibnamefont  [1]{#1}%
\providecommand \bibfnamefont [1]{#1}%
\providecommand \citenamefont [1]{#1}%
\providecommand \href@noop [0]{\@secondoftwo}%
\providecommand \href [0]{\begingroup \@sanitize@url \@href}%
\providecommand \@href[1]{\@@startlink{#1}\@@href}%
\providecommand \@@href[1]{\endgroup#1\@@endlink}%
\providecommand \@sanitize@url [0]{\catcode `\\12\catcode `\$12\catcode
  `\&12\catcode `\#12\catcode `\^12\catcode `\_12\catcode `\%12\relax}%
\providecommand \@@startlink[1]{}%
\providecommand \@@endlink[0]{}%
\providecommand \url  [0]{\begingroup\@sanitize@url \@url }%
\providecommand \@url [1]{\endgroup\@href {#1}{\urlprefix }}%
\providecommand \urlprefix  [0]{URL }%
\providecommand \Eprint [0]{\href }%
\providecommand \doibase [0]{http://dx.doi.org/}%
\providecommand \selectlanguage [0]{\@gobble}%
\providecommand \bibinfo  [0]{\@secondoftwo}%
\providecommand \bibfield  [0]{\@secondoftwo}%
\providecommand \translation [1]{[#1]}%
\providecommand \BibitemOpen [0]{}%
\providecommand \bibitemStop [0]{}%
\providecommand \bibitemNoStop [0]{.\EOS\space}%
\providecommand \EOS [0]{\spacefactor3000\relax}%
\providecommand \BibitemShut  [1]{\csname bibitem#1\endcsname}%
\let\auto@bib@innerbib\@empty
\bibitem [{\citenamefont {Levine}\ \emph {et~al.}(2019)\citenamefont {Levine},
  \citenamefont {Turner}, \citenamefont {Kehayias}, \citenamefont {Hart},
  \citenamefont {Langellier}, \citenamefont {Trubko}, \citenamefont {Glenn},
  \citenamefont {Fu},\ and\ \citenamefont {Walsworth}}]{levine2019principles}%
  \BibitemOpen
  \bibfield  {author} {\bibinfo {author} {\bibfnamefont {E.~V.}\ \bibnamefont
  {Levine}}, \bibinfo {author} {\bibfnamefont {M.~J.}\ \bibnamefont {Turner}},
  \bibinfo {author} {\bibfnamefont {P.}~\bibnamefont {Kehayias}}, \bibinfo
  {author} {\bibfnamefont {C.~A.}\ \bibnamefont {Hart}}, \bibinfo {author}
  {\bibfnamefont {N.}~\bibnamefont {Langellier}}, \bibinfo {author}
  {\bibfnamefont {R.}~\bibnamefont {Trubko}}, \bibinfo {author} {\bibfnamefont
  {D.~R.}\ \bibnamefont {Glenn}}, \bibinfo {author} {\bibfnamefont {R.~R.}\
  \bibnamefont {Fu}}, \ and\ \bibinfo {author} {\bibfnamefont {R.~L.}\
  \bibnamefont {Walsworth}},\ }\bibfield  {title} {\enquote {\bibinfo {title}
  {Principles and techniques of the quantum diamond microscope},}\ }\href
  {\doibase https://doi.org/10.1515/nanoph-2019-0209} {\bibfield  {journal}
  {\bibinfo  {journal} {Nanophotonics}\ }\textbf {\bibinfo {volume} {8}},\
  \bibinfo {pages} {1945--1973} (\bibinfo {year} {2019})}\BibitemShut {NoStop}%
\bibitem [{\citenamefont {Barry}\ \emph {et~al.}(2020)\citenamefont {Barry},
  \citenamefont {Schloss}, \citenamefont {Bauch}, \citenamefont {Turner},
  \citenamefont {Hart}, \citenamefont {Pham},\ and\ \citenamefont
  {Walsworth}}]{barry2020report}%
  \BibitemOpen
  \bibfield  {author} {\bibinfo {author} {\bibfnamefont {J.~F.}\ \bibnamefont
  {Barry}}, \bibinfo {author} {\bibfnamefont {J.~M.}\ \bibnamefont {Schloss}},
  \bibinfo {author} {\bibfnamefont {E.}~\bibnamefont {Bauch}}, \bibinfo
  {author} {\bibfnamefont {M.~J.}\ \bibnamefont {Turner}}, \bibinfo {author}
  {\bibfnamefont {C.~A.}\ \bibnamefont {Hart}}, \bibinfo {author}
  {\bibfnamefont {L.~M.}\ \bibnamefont {Pham}}, \ and\ \bibinfo {author}
  {\bibfnamefont {R.~L.}\ \bibnamefont {Walsworth}},\ }\bibfield  {title}
  {\enquote {\bibinfo {title} {Sensitivity optimization for {NV}-diamond
  magnetometry},}\ }\href {\doibase 10.1103/RevModPhys.92.015004} {\bibfield
  {journal} {\bibinfo  {journal} {Rev. Mod. Phys.}\ }\textbf {\bibinfo {volume}
  {92}},\ \bibinfo {pages} {015004} (\bibinfo {year} {2020})}\BibitemShut
  {NoStop}%
\bibitem [{\citenamefont {Aslam}\ \emph {et~al.}(2023)\citenamefont {Aslam},
  \citenamefont {Zhou}, \citenamefont {Urbach}, \citenamefont {Turner},
  \citenamefont {Walsworth}, \citenamefont {Lukin},\ and\ \citenamefont
  {Park}}]{aslam2023quantum}%
  \BibitemOpen
  \bibfield  {author} {\bibinfo {author} {\bibfnamefont {N.}~\bibnamefont
  {Aslam}}, \bibinfo {author} {\bibfnamefont {H.}~\bibnamefont {Zhou}},
  \bibinfo {author} {\bibfnamefont {E.~K.}\ \bibnamefont {Urbach}}, \bibinfo
  {author} {\bibfnamefont {M.~J.}\ \bibnamefont {Turner}}, \bibinfo {author}
  {\bibfnamefont {R.~L.}\ \bibnamefont {Walsworth}}, \bibinfo {author}
  {\bibfnamefont {M.~D.}\ \bibnamefont {Lukin}}, \ and\ \bibinfo {author}
  {\bibfnamefont {H.}~\bibnamefont {Park}},\ }\bibfield  {title} {\enquote
  {\bibinfo {title} {Quantum sensors for biomedical applications},}\ }\href
  {\doibase 10.1038/s42254-023-00558-3} {\bibfield  {journal} {\bibinfo
  {journal} {Nat. Rev. Phys.}\ }\textbf {\bibinfo {volume} {5}},\ \bibinfo
  {pages} {157--169} (\bibinfo {year} {2023})}\BibitemShut {NoStop}%
\bibitem [{\citenamefont {Glenn}\ \emph {et~al.}(2017)\citenamefont {Glenn},
  \citenamefont {Fu}, \citenamefont {Kehayias}, \citenamefont {Le~Sage},
  \citenamefont {Lima}, \citenamefont {Weiss},\ and\ \citenamefont
  {Walsworth}}]{glenn2017micrometer}%
  \BibitemOpen
  \bibfield  {author} {\bibinfo {author} {\bibfnamefont {D.~R.}\ \bibnamefont
  {Glenn}}, \bibinfo {author} {\bibfnamefont {R.~R.}\ \bibnamefont {Fu}},
  \bibinfo {author} {\bibfnamefont {P.}~\bibnamefont {Kehayias}}, \bibinfo
  {author} {\bibfnamefont {D.}~\bibnamefont {Le~Sage}}, \bibinfo {author}
  {\bibfnamefont {E.~A.}\ \bibnamefont {Lima}}, \bibinfo {author}
  {\bibfnamefont {B.~P.}\ \bibnamefont {Weiss}}, \ and\ \bibinfo {author}
  {\bibfnamefont {R.~L.}\ \bibnamefont {Walsworth}},\ }\bibfield  {title}
  {\enquote {\bibinfo {title} {Micrometer-scale magnetic imaging of geological
  samples using a quantum diamond microscope},}\ }\href {\doibase
  10.1002/2017GC006946} {\bibfield  {journal} {\bibinfo  {journal} {Geochem.
  Geophys. Geosyst.}\ }\textbf {\bibinfo {volume} {18}},\ \bibinfo {pages}
  {3254--3267} (\bibinfo {year} {2017})}\BibitemShut {NoStop}%
\bibitem [{\citenamefont {Ku}\ \emph {et~al.}(2020)\citenamefont {Ku},
  \citenamefont {Zhou}, \citenamefont {Li}, \citenamefont {Shin}, \citenamefont
  {Shi}, \citenamefont {Burch}, \citenamefont {Anderson}, \citenamefont
  {Pierce}, \citenamefont {Xie}, \citenamefont {Hamo} \emph
  {et~al.}}]{ku2020imaging}%
  \BibitemOpen
  \bibfield  {author} {\bibinfo {author} {\bibfnamefont {M.~J.}\ \bibnamefont
  {Ku}}, \bibinfo {author} {\bibfnamefont {T.~X.}\ \bibnamefont {Zhou}},
  \bibinfo {author} {\bibfnamefont {Q.}~\bibnamefont {Li}}, \bibinfo {author}
  {\bibfnamefont {Y.~J.}\ \bibnamefont {Shin}}, \bibinfo {author}
  {\bibfnamefont {J.~K.}\ \bibnamefont {Shi}}, \bibinfo {author} {\bibfnamefont
  {C.}~\bibnamefont {Burch}}, \bibinfo {author} {\bibfnamefont {L.~E.}\
  \bibnamefont {Anderson}}, \bibinfo {author} {\bibfnamefont {A.~T.}\
  \bibnamefont {Pierce}}, \bibinfo {author} {\bibfnamefont {Y.}~\bibnamefont
  {Xie}}, \bibinfo {author} {\bibfnamefont {A.}~\bibnamefont {Hamo}},  \emph
  {et~al.},\ }\bibfield  {title} {\enquote {\bibinfo {title} {Imaging viscous
  flow of the dirac fluid in graphene},}\ }\href {\doibase
  https://doi.org/10.1038/s41586-020-2507-2} {\bibfield  {journal} {\bibinfo
  {journal} {Nature}\ }\textbf {\bibinfo {volume} {583}},\ \bibinfo {pages}
  {537--541} (\bibinfo {year} {2020})}\BibitemShut {NoStop}%
\bibitem [{\citenamefont {Turner}\ \emph {et~al.}(2020)\citenamefont {Turner},
  \citenamefont {Langellier}, \citenamefont {Bainbridge}, \citenamefont
  {Walters}, \citenamefont {Meesala}, \citenamefont {Babinec}, \citenamefont
  {Kehayias}, \citenamefont {Yacoby}, \citenamefont {Hu}, \citenamefont
  {Lon\ifmmode~\check{c}\else \v{c}\fi{}ar}, \citenamefont {Walsworth},\ and\
  \citenamefont {Levine}}]{Turner2020IC}%
  \BibitemOpen
  \bibfield  {author} {\bibinfo {author} {\bibfnamefont {M.~J.}\ \bibnamefont
  {Turner}}, \bibinfo {author} {\bibfnamefont {N.}~\bibnamefont {Langellier}},
  \bibinfo {author} {\bibfnamefont {R.}~\bibnamefont {Bainbridge}}, \bibinfo
  {author} {\bibfnamefont {D.}~\bibnamefont {Walters}}, \bibinfo {author}
  {\bibfnamefont {S.}~\bibnamefont {Meesala}}, \bibinfo {author} {\bibfnamefont
  {T.~M.}\ \bibnamefont {Babinec}}, \bibinfo {author} {\bibfnamefont
  {P.}~\bibnamefont {Kehayias}}, \bibinfo {author} {\bibfnamefont
  {A.}~\bibnamefont {Yacoby}}, \bibinfo {author} {\bibfnamefont
  {E.}~\bibnamefont {Hu}}, \bibinfo {author} {\bibfnamefont {M.}~\bibnamefont
  {Lon\ifmmode~\check{c}\else \v{c}\fi{}ar}}, \bibinfo {author} {\bibfnamefont
  {R.~L.}\ \bibnamefont {Walsworth}}, \ and\ \bibinfo {author} {\bibfnamefont
  {E.~V.}\ \bibnamefont {Levine}},\ }\bibfield  {title} {\enquote {\bibinfo
  {title} {Magnetic field fingerprinting of integrated-circuit activity with a
  quantum diamond microscope},}\ }\href {\doibase
  10.1103/PhysRevApplied.14.014097} {\bibfield  {journal} {\bibinfo  {journal}
  {Phys. Rev. Appl.}\ }\textbf {\bibinfo {volume} {14}},\ \bibinfo {pages}
  {014097} (\bibinfo {year} {2020})}\BibitemShut {NoStop}%
\bibitem [{\citenamefont {Le~Sage}\ \emph {et~al.}(2013)\citenamefont
  {Le~Sage}, \citenamefont {Arai}, \citenamefont {Glenn}, \citenamefont
  {DeVience}, \citenamefont {Pham}, \citenamefont {Rahn-Lee}, \citenamefont
  {Lukin}, \citenamefont {Yacoby}, \citenamefont {Komeili},\ and\ \citenamefont
  {Walsworth}}]{le2013optical}%
  \BibitemOpen
  \bibfield  {author} {\bibinfo {author} {\bibfnamefont {D.}~\bibnamefont
  {Le~Sage}}, \bibinfo {author} {\bibfnamefont {K.}~\bibnamefont {Arai}},
  \bibinfo {author} {\bibfnamefont {D.~R.}\ \bibnamefont {Glenn}}, \bibinfo
  {author} {\bibfnamefont {S.~J.}\ \bibnamefont {DeVience}}, \bibinfo {author}
  {\bibfnamefont {L.~M.}\ \bibnamefont {Pham}}, \bibinfo {author}
  {\bibfnamefont {L.}~\bibnamefont {Rahn-Lee}}, \bibinfo {author}
  {\bibfnamefont {M.~D.}\ \bibnamefont {Lukin}}, \bibinfo {author}
  {\bibfnamefont {A.}~\bibnamefont {Yacoby}}, \bibinfo {author} {\bibfnamefont
  {A.}~\bibnamefont {Komeili}}, \ and\ \bibinfo {author} {\bibfnamefont
  {R.~L.}\ \bibnamefont {Walsworth}},\ }\bibfield  {title} {\enquote {\bibinfo
  {title} {Optical magnetic imaging of living cells},}\ }\href {\doibase
  10.1038/nature12072} {\bibfield  {journal} {\bibinfo  {journal} {Nature}\
  }\textbf {\bibinfo {volume} {496}},\ \bibinfo {pages} {486--489} (\bibinfo
  {year} {2013})}\BibitemShut {NoStop}%
\bibitem [{\citenamefont {Fescenko}\ \emph {et~al.}(2019)\citenamefont
  {Fescenko}, \citenamefont {Laraoui}, \citenamefont {Smits}, \citenamefont
  {Mosavian}, \citenamefont {Kehayias}, \citenamefont {Seto}, \citenamefont
  {Bougas}, \citenamefont {Jarmola},\ and\ \citenamefont
  {Acosta}}]{fescenko2019diamond}%
  \BibitemOpen
  \bibfield  {author} {\bibinfo {author} {\bibfnamefont {I.}~\bibnamefont
  {Fescenko}}, \bibinfo {author} {\bibfnamefont {A.}~\bibnamefont {Laraoui}},
  \bibinfo {author} {\bibfnamefont {J.}~\bibnamefont {Smits}}, \bibinfo
  {author} {\bibfnamefont {N.}~\bibnamefont {Mosavian}}, \bibinfo {author}
  {\bibfnamefont {P.}~\bibnamefont {Kehayias}}, \bibinfo {author}
  {\bibfnamefont {J.}~\bibnamefont {Seto}}, \bibinfo {author} {\bibfnamefont
  {L.}~\bibnamefont {Bougas}}, \bibinfo {author} {\bibfnamefont
  {A.}~\bibnamefont {Jarmola}}, \ and\ \bibinfo {author} {\bibfnamefont
  {V.~M.}\ \bibnamefont {Acosta}},\ }\bibfield  {title} {\enquote {\bibinfo
  {title} {Diamond magnetic microscopy of malarial hemozoin nanocrystals},}\
  }\href {\doibase 10.1103/PhysRevApplied.11.034029} {\bibfield  {journal}
  {\bibinfo  {journal} {Phys. Rev. Appl.}\ }\textbf {\bibinfo {volume} {11}},\
  \bibinfo {pages} {034029} (\bibinfo {year} {2019})}\BibitemShut {NoStop}%
\bibitem [{\citenamefont {de~Gille}\ \emph {et~al.}(2021)\citenamefont
  {de~Gille}, \citenamefont {McCoey}, \citenamefont {Hall}, \citenamefont
  {Tetienne}, \citenamefont {Malkemper}, \citenamefont {Keays}, \citenamefont
  {Hollenberg},\ and\ \citenamefont {Simpson}}]{de2021quantum}%
  \BibitemOpen
  \bibfield  {author} {\bibinfo {author} {\bibfnamefont {R.~W.}\ \bibnamefont
  {de~Gille}}, \bibinfo {author} {\bibfnamefont {J.~M.}\ \bibnamefont
  {McCoey}}, \bibinfo {author} {\bibfnamefont {L.~T.}\ \bibnamefont {Hall}},
  \bibinfo {author} {\bibfnamefont {J.-P.}\ \bibnamefont {Tetienne}}, \bibinfo
  {author} {\bibfnamefont {E.~P.}\ \bibnamefont {Malkemper}}, \bibinfo {author}
  {\bibfnamefont {D.~A.}\ \bibnamefont {Keays}}, \bibinfo {author}
  {\bibfnamefont {L.~C.}\ \bibnamefont {Hollenberg}}, \ and\ \bibinfo {author}
  {\bibfnamefont {D.~A.}\ \bibnamefont {Simpson}},\ }\bibfield  {title}
  {\enquote {\bibinfo {title} {Quantum magnetic imaging of iron organelles
  within the pigeon cochlea},}\ }\href {\doibase 10.1073/pnas.2112749118}
  {\bibfield  {journal} {\bibinfo  {journal} {Proc. Natl. Acad. Sci. U.S.A.}\
  }\textbf {\bibinfo {volume} {118}},\ \bibinfo {pages} {e2112749118} (\bibinfo
  {year} {2021})}\BibitemShut {NoStop}%
\bibitem [{\citenamefont {Davis}\ \emph {et~al.}(2018)\citenamefont {Davis},
  \citenamefont {Ramesh}, \citenamefont {Bhatnagar}, \citenamefont
  {Lee-Gosselin}, \citenamefont {Barry}, \citenamefont {Glenn}, \citenamefont
  {Walsworth},\ and\ \citenamefont {Shapiro}}]{davis2018mapping}%
  \BibitemOpen
  \bibfield  {author} {\bibinfo {author} {\bibfnamefont {H.~C.}\ \bibnamefont
  {Davis}}, \bibinfo {author} {\bibfnamefont {P.}~\bibnamefont {Ramesh}},
  \bibinfo {author} {\bibfnamefont {A.}~\bibnamefont {Bhatnagar}}, \bibinfo
  {author} {\bibfnamefont {A.}~\bibnamefont {Lee-Gosselin}}, \bibinfo {author}
  {\bibfnamefont {J.~F.}\ \bibnamefont {Barry}}, \bibinfo {author}
  {\bibfnamefont {D.~R.}\ \bibnamefont {Glenn}}, \bibinfo {author}
  {\bibfnamefont {R.~L.}\ \bibnamefont {Walsworth}}, \ and\ \bibinfo {author}
  {\bibfnamefont {M.~G.}\ \bibnamefont {Shapiro}},\ }\bibfield  {title}
  {\enquote {\bibinfo {title} {Mapping the microscale origins of magnetic
  resonance image contrast with subcellular diamond magnetometry},}\ }\href
  {\doibase https://doi.org/10.1038/s41467-017-02471-7} {\bibfield  {journal}
  {\bibinfo  {journal} {Nat. Commun.}\ }\textbf {\bibinfo {volume} {9}},\
  \bibinfo {pages} {131} (\bibinfo {year} {2018})}\BibitemShut {NoStop}%
\bibitem [{\citenamefont {Kazi}\ \emph {et~al.}(2021)\citenamefont {Kazi},
  \citenamefont {Shelby}, \citenamefont {Watanabe}, \citenamefont {Itoh},
  \citenamefont {Shutthanandan}, \citenamefont {Wiggins},\ and\ \citenamefont
  {Fu}}]{Kazi2021wide}%
  \BibitemOpen
  \bibfield  {author} {\bibinfo {author} {\bibfnamefont {Z.}~\bibnamefont
  {Kazi}}, \bibinfo {author} {\bibfnamefont {I.~M.}\ \bibnamefont {Shelby}},
  \bibinfo {author} {\bibfnamefont {H.}~\bibnamefont {Watanabe}}, \bibinfo
  {author} {\bibfnamefont {K.~M.}\ \bibnamefont {Itoh}}, \bibinfo {author}
  {\bibfnamefont {V.}~\bibnamefont {Shutthanandan}}, \bibinfo {author}
  {\bibfnamefont {P.~A.}\ \bibnamefont {Wiggins}}, \ and\ \bibinfo {author}
  {\bibfnamefont {K.-M.~C.}\ \bibnamefont {Fu}},\ }\bibfield  {title} {\enquote
  {\bibinfo {title} {Wide-field dynamic magnetic microscopy using double-double
  quantum driving of a diamond defect ensemble},}\ }\href {\doibase
  10.1103/PhysRevApplied.15.054032} {\bibfield  {journal} {\bibinfo  {journal}
  {Phys. Rev. Appl.}\ }\textbf {\bibinfo {volume} {15}},\ \bibinfo {pages}
  {054032} (\bibinfo {year} {2021})}\BibitemShut {NoStop}%
\bibitem [{\citenamefont {Glenn}\ \emph {et~al.}(2015)\citenamefont {Glenn},
  \citenamefont {Lee}, \citenamefont {Park}, \citenamefont {Weissleder},
  \citenamefont {Yacoby}, \citenamefont {Lukin}, \citenamefont {Lee},
  \citenamefont {Walsworth},\ and\ \citenamefont {Connolly}}]{glenn2015single}%
  \BibitemOpen
  \bibfield  {author} {\bibinfo {author} {\bibfnamefont {D.~R.}\ \bibnamefont
  {Glenn}}, \bibinfo {author} {\bibfnamefont {K.}~\bibnamefont {Lee}}, \bibinfo
  {author} {\bibfnamefont {H.}~\bibnamefont {Park}}, \bibinfo {author}
  {\bibfnamefont {R.}~\bibnamefont {Weissleder}}, \bibinfo {author}
  {\bibfnamefont {A.}~\bibnamefont {Yacoby}}, \bibinfo {author} {\bibfnamefont
  {M.~D.}\ \bibnamefont {Lukin}}, \bibinfo {author} {\bibfnamefont
  {H.}~\bibnamefont {Lee}}, \bibinfo {author} {\bibfnamefont {R.~L.}\
  \bibnamefont {Walsworth}}, \ and\ \bibinfo {author} {\bibfnamefont {C.~B.}\
  \bibnamefont {Connolly}},\ }\bibfield  {title} {\enquote {\bibinfo {title}
  {Single-cell magnetic imaging using a quantum diamond microscope},}\ }\href
  {\doibase 10.1038/nmeth.3449} {\bibfield  {journal} {\bibinfo  {journal}
  {Nat. Methods}\ }\textbf {\bibinfo {volume} {12}},\ \bibinfo {pages}
  {736--738} (\bibinfo {year} {2015})}\BibitemShut {NoStop}%
\bibitem [{\citenamefont {McCoey}\ \emph {et~al.}(2020)\citenamefont {McCoey},
  \citenamefont {Matsuoka}, \citenamefont {de~Gille}, \citenamefont {Hall},
  \citenamefont {Shaw}, \citenamefont {Tetienne}, \citenamefont {Kisailus},
  \citenamefont {Hollenberg},\ and\ \citenamefont
  {Simpson}}]{mccoey2020quantum}%
  \BibitemOpen
  \bibfield  {author} {\bibinfo {author} {\bibfnamefont {J.~M.}\ \bibnamefont
  {McCoey}}, \bibinfo {author} {\bibfnamefont {M.}~\bibnamefont {Matsuoka}},
  \bibinfo {author} {\bibfnamefont {R.~W.}\ \bibnamefont {de~Gille}}, \bibinfo
  {author} {\bibfnamefont {L.~T.}\ \bibnamefont {Hall}}, \bibinfo {author}
  {\bibfnamefont {J.~A.}\ \bibnamefont {Shaw}}, \bibinfo {author}
  {\bibfnamefont {J.-P.}\ \bibnamefont {Tetienne}}, \bibinfo {author}
  {\bibfnamefont {D.}~\bibnamefont {Kisailus}}, \bibinfo {author}
  {\bibfnamefont {L.~C.}\ \bibnamefont {Hollenberg}}, \ and\ \bibinfo {author}
  {\bibfnamefont {D.~A.}\ \bibnamefont {Simpson}},\ }\bibfield  {title}
  {\enquote {\bibinfo {title} {Quantum magnetic imaging of iron
  biomineralization in teeth of the chiton {Acanthopleura} hirtosa},}\ }\href
  {\doibase https://doi.org/10.1002/smtd.201900754} {\bibfield  {journal}
  {\bibinfo  {journal} {Small Methods}\ }\textbf {\bibinfo {volume} {4}},\
  \bibinfo {pages} {1900754} (\bibinfo {year} {2020})}\BibitemShut {NoStop}%
\bibitem [{\citenamefont {Troise}\ \emph {et~al.}(2022)\citenamefont {Troise},
  \citenamefont {Hansen}, \citenamefont {Olsson}, \citenamefont {Webb},
  \citenamefont {Tomasevic}, \citenamefont {Achard}, \citenamefont {Brinza},
  \citenamefont {Staacke}, \citenamefont {Kieschnick}, \citenamefont {Meijer}
  \emph {et~al.}}]{troise2022vitro}%
  \BibitemOpen
  \bibfield  {author} {\bibinfo {author} {\bibfnamefont {L.}~\bibnamefont
  {Troise}}, \bibinfo {author} {\bibfnamefont {N.~W.}\ \bibnamefont {Hansen}},
  \bibinfo {author} {\bibfnamefont {C.}~\bibnamefont {Olsson}}, \bibinfo
  {author} {\bibfnamefont {J.~L.}\ \bibnamefont {Webb}}, \bibinfo {author}
  {\bibfnamefont {L.}~\bibnamefont {Tomasevic}}, \bibinfo {author}
  {\bibfnamefont {J.}~\bibnamefont {Achard}}, \bibinfo {author} {\bibfnamefont
  {O.}~\bibnamefont {Brinza}}, \bibinfo {author} {\bibfnamefont
  {R.}~\bibnamefont {Staacke}}, \bibinfo {author} {\bibfnamefont
  {M.}~\bibnamefont {Kieschnick}}, \bibinfo {author} {\bibfnamefont
  {J.}~\bibnamefont {Meijer}},  \emph {et~al.},\ }\bibfield  {title} {\enquote
  {\bibinfo {title} {In vitro recording of muscle activity induced by high
  intensity laser optogenetic stimulation using a diamond quantum biosensor},}\
  }\href {\doibase https://doi.org/10.1116/5.0106099} {\bibfield  {journal}
  {\bibinfo  {journal} {AVS Quantum Science}\ }\textbf {\bibinfo {volume}
  {4}},\ \bibinfo {pages} {044402} (\bibinfo {year} {2022})}\BibitemShut
  {NoStop}%
\bibitem [{\citenamefont {Barry}\ \emph {et~al.}(2016)\citenamefont {Barry},
  \citenamefont {Turner}, \citenamefont {Schloss}, \citenamefont {Glenn},
  \citenamefont {Song}, \citenamefont {Lukin}, \citenamefont {Park},\ and\
  \citenamefont {Walsworth}}]{barry2016optical}%
  \BibitemOpen
  \bibfield  {author} {\bibinfo {author} {\bibfnamefont {J.~F.}\ \bibnamefont
  {Barry}}, \bibinfo {author} {\bibfnamefont {M.~J.}\ \bibnamefont {Turner}},
  \bibinfo {author} {\bibfnamefont {J.~M.}\ \bibnamefont {Schloss}}, \bibinfo
  {author} {\bibfnamefont {D.~R.}\ \bibnamefont {Glenn}}, \bibinfo {author}
  {\bibfnamefont {Y.}~\bibnamefont {Song}}, \bibinfo {author} {\bibfnamefont
  {M.~D.}\ \bibnamefont {Lukin}}, \bibinfo {author} {\bibfnamefont
  {H.}~\bibnamefont {Park}}, \ and\ \bibinfo {author} {\bibfnamefont {R.~L.}\
  \bibnamefont {Walsworth}},\ }\bibfield  {title} {\enquote {\bibinfo {title}
  {Optical magnetic detection of single-neuron action potentials using quantum
  defects in diamond},}\ }\href {\doibase 10.1073/pnas.1601513113} {\bibfield
  {journal} {\bibinfo  {journal} {Proc. Natl. Acad. Sci. U.S.A.}\ }\textbf
  {\bibinfo {volume} {113}},\ \bibinfo {pages} {14133--14138} (\bibinfo {year}
  {2016})}\BibitemShut {NoStop}%
\bibitem [{\citenamefont {Arai}\ \emph {et~al.}(2022)\citenamefont {Arai},
  \citenamefont {Kuwahata}, \citenamefont {Nishitani}, \citenamefont
  {Fujisaki}, \citenamefont {Matsuki}, \citenamefont {Nishio}, \citenamefont
  {Xin}, \citenamefont {Cao}, \citenamefont {Hatano}, \citenamefont {Onoda}
  \emph {et~al.}}]{arai2022millimetre}%
  \BibitemOpen
  \bibfield  {author} {\bibinfo {author} {\bibfnamefont {K.}~\bibnamefont
  {Arai}}, \bibinfo {author} {\bibfnamefont {A.}~\bibnamefont {Kuwahata}},
  \bibinfo {author} {\bibfnamefont {D.}~\bibnamefont {Nishitani}}, \bibinfo
  {author} {\bibfnamefont {I.}~\bibnamefont {Fujisaki}}, \bibinfo {author}
  {\bibfnamefont {R.}~\bibnamefont {Matsuki}}, \bibinfo {author} {\bibfnamefont
  {Y.}~\bibnamefont {Nishio}}, \bibinfo {author} {\bibfnamefont
  {Z.}~\bibnamefont {Xin}}, \bibinfo {author} {\bibfnamefont {X.}~\bibnamefont
  {Cao}}, \bibinfo {author} {\bibfnamefont {Y.}~\bibnamefont {Hatano}},
  \bibinfo {author} {\bibfnamefont {S.}~\bibnamefont {Onoda}},  \emph
  {et~al.},\ }\bibfield  {title} {\enquote {\bibinfo {title} {Millimetre-scale
  magnetocardiography of living rats with thoracotomy},}\ }\href {\doibase
  10.1038/s42005-022-00978-0} {\bibfield  {journal} {\bibinfo  {journal}
  {Commun. Phys.}\ }\textbf {\bibinfo {volume} {5}},\ \bibinfo {pages} {200}
  (\bibinfo {year} {2022})}\BibitemShut {NoStop}%
\bibitem [{\citenamefont {Hart}\ \emph {et~al.}(2021)\citenamefont {Hart},
  \citenamefont {Schloss}, \citenamefont {Turner}, \citenamefont {Scheidegger},
  \citenamefont {Bauch},\ and\ \citenamefont {Walsworth}}]{Hart2021-4Ramsey}%
  \BibitemOpen
  \bibfield  {author} {\bibinfo {author} {\bibfnamefont {C.~A.}\ \bibnamefont
  {Hart}}, \bibinfo {author} {\bibfnamefont {J.~M.}\ \bibnamefont {Schloss}},
  \bibinfo {author} {\bibfnamefont {M.~J.}\ \bibnamefont {Turner}}, \bibinfo
  {author} {\bibfnamefont {P.~J.}\ \bibnamefont {Scheidegger}}, \bibinfo
  {author} {\bibfnamefont {E.}~\bibnamefont {Bauch}}, \ and\ \bibinfo {author}
  {\bibfnamefont {R.~L.}\ \bibnamefont {Walsworth}},\ }\bibfield  {title}
  {\enquote {\bibinfo {title} {$\mathrm{N}$-{$V$}--diamond magnetic microscopy
  using a double quantum 4-ramsey protocol},}\ }\href {\doibase
  10.1103/PhysRevApplied.15.044020} {\bibfield  {journal} {\bibinfo  {journal}
  {Phys. Rev. Appl.}\ }\textbf {\bibinfo {volume} {15}},\ \bibinfo {pages}
  {044020} (\bibinfo {year} {2021})}\BibitemShut {NoStop}%
\bibitem [{\citenamefont {Turner}(2020)}]{turner2020huphd}%
  \BibitemOpen
  \bibfield  {author} {\bibinfo {author} {\bibfnamefont {M.~J.}\ \bibnamefont
  {Turner}},\ }\emph {\bibinfo {title} {Quantum diamond microscopes for
  biological systems and integrated circuits}},\ \href@noop {} {Ph.D. thesis},\
  \bibinfo  {school} {Harvard University} (\bibinfo {year} {2020})\BibitemShut
  {NoStop}%
\bibitem [{\citenamefont {Schloss}(2019)}]{schloss2019optimizing}%
  \BibitemOpen
  \bibfield  {author} {\bibinfo {author} {\bibfnamefont {J.~M.}\ \bibnamefont
  {Schloss}},\ }\emph {\bibinfo {title} {Optimizing nitrogen-vacancy diamond
  magnetic sensors and imagers for broadband sensitivity}},\ \href@noop {}
  {Ph.D. thesis},\ \bibinfo  {school} {Massachusetts Institute of Technology}
  (\bibinfo {year} {2019})\BibitemShut {NoStop}%
\bibitem [{\citenamefont {Bauch}\ \emph {et~al.}(2018)\citenamefont {Bauch},
  \citenamefont {Hart}, \citenamefont {Schloss}, \citenamefont {Turner},
  \citenamefont {Barry}, \citenamefont {Kehayias}, \citenamefont {Singh},\ and\
  \citenamefont {Walsworth}}]{bauch2018ultralong}%
  \BibitemOpen
  \bibfield  {author} {\bibinfo {author} {\bibfnamefont {E.}~\bibnamefont
  {Bauch}}, \bibinfo {author} {\bibfnamefont {C.~A.}\ \bibnamefont {Hart}},
  \bibinfo {author} {\bibfnamefont {J.~M.}\ \bibnamefont {Schloss}}, \bibinfo
  {author} {\bibfnamefont {M.~J.}\ \bibnamefont {Turner}}, \bibinfo {author}
  {\bibfnamefont {J.~F.}\ \bibnamefont {Barry}}, \bibinfo {author}
  {\bibfnamefont {P.}~\bibnamefont {Kehayias}}, \bibinfo {author}
  {\bibfnamefont {S.}~\bibnamefont {Singh}}, \ and\ \bibinfo {author}
  {\bibfnamefont {R.~L.}\ \bibnamefont {Walsworth}},\ }\bibfield  {title}
  {\enquote {\bibinfo {title} {Ultralong dephasing times in solid-state spin
  ensembles via quantum control},}\ }\href {\doibase 10.1103/PhysRevX.8.031025}
  {\bibfield  {journal} {\bibinfo  {journal} {Phys. Rev. X}\ }\textbf {\bibinfo
  {volume} {8}},\ \bibinfo {pages} {031025} (\bibinfo {year}
  {2018})}\BibitemShut {NoStop}%
\bibitem [{\citenamefont {Buades}, \citenamefont {Coll},\ and\ \citenamefont
  {Morel}(2011)}]{buades2011non}%
  \BibitemOpen
  \bibfield  {author} {\bibinfo {author} {\bibfnamefont {A.}~\bibnamefont
  {Buades}}, \bibinfo {author} {\bibfnamefont {B.}~\bibnamefont {Coll}}, \ and\
  \bibinfo {author} {\bibfnamefont {J.-M.}\ \bibnamefont {Morel}},\ }\bibfield
  {title} {\enquote {\bibinfo {title} {Non-local means denoising},}\ }\href
  {\doibase https://doi.org/10.5201/ipol.2011.bcm_nlm} {\bibfield  {journal}
  {\bibinfo  {journal} {Image Process. Line}\ }\textbf {\bibinfo {volume}
  {1}},\ \bibinfo {pages} {208--212} (\bibinfo {year} {2011})}\BibitemShut
  {NoStop}%
\bibitem [{\citenamefont {Sigmund}\ \emph {et~al.}(2018)\citenamefont
  {Sigmund}, \citenamefont {Massner}, \citenamefont {Erdmann}, \citenamefont
  {Stelzl}, \citenamefont {Rolbieski}, \citenamefont {Desai}, \citenamefont
  {Bricault}, \citenamefont {W{\"o}rner}, \citenamefont {Snijder},
  \citenamefont {Geerlof} \emph {et~al.}}]{sigmund2018bacterial}%
  \BibitemOpen
  \bibfield  {author} {\bibinfo {author} {\bibfnamefont {F.}~\bibnamefont
  {Sigmund}}, \bibinfo {author} {\bibfnamefont {C.}~\bibnamefont {Massner}},
  \bibinfo {author} {\bibfnamefont {P.}~\bibnamefont {Erdmann}}, \bibinfo
  {author} {\bibfnamefont {A.}~\bibnamefont {Stelzl}}, \bibinfo {author}
  {\bibfnamefont {H.}~\bibnamefont {Rolbieski}}, \bibinfo {author}
  {\bibfnamefont {M.}~\bibnamefont {Desai}}, \bibinfo {author} {\bibfnamefont
  {S.}~\bibnamefont {Bricault}}, \bibinfo {author} {\bibfnamefont {T.~P.}\
  \bibnamefont {W{\"o}rner}}, \bibinfo {author} {\bibfnamefont
  {J.}~\bibnamefont {Snijder}}, \bibinfo {author} {\bibfnamefont
  {A.}~\bibnamefont {Geerlof}},  \emph {et~al.},\ }\bibfield  {title} {\enquote
  {\bibinfo {title} {Bacterial encapsulins as orthogonal compartments for
  mammalian cell engineering},}\ }\href {\doibase
  https://doi.org/10.1038/s41467-018-04227-3} {\bibfield  {journal} {\bibinfo
  {journal} {Nat. Commun.}\ }\textbf {\bibinfo {volume} {9}},\ \bibinfo {pages}
  {1990} (\bibinfo {year} {2018})}\BibitemShut {NoStop}%
\bibitem [{\citenamefont {Diebel}\ \emph {et~al.}(2000)\citenamefont {Diebel},
  \citenamefont {Proksch}, \citenamefont {Green}, \citenamefont {Neilson},\
  and\ \citenamefont {Walker}}]{diebel2000magnetite}%
  \BibitemOpen
  \bibfield  {author} {\bibinfo {author} {\bibfnamefont {C.~E.}\ \bibnamefont
  {Diebel}}, \bibinfo {author} {\bibfnamefont {R.}~\bibnamefont {Proksch}},
  \bibinfo {author} {\bibfnamefont {C.~R.}\ \bibnamefont {Green}}, \bibinfo
  {author} {\bibfnamefont {P.}~\bibnamefont {Neilson}}, \ and\ \bibinfo
  {author} {\bibfnamefont {M.~M.}\ \bibnamefont {Walker}},\ }\bibfield  {title}
  {\enquote {\bibinfo {title} {Magnetite defines a vertebrate
  magnetoreceptor},}\ }\href {\doibase 10.1038/35018561} {\bibfield  {journal}
  {\bibinfo  {journal} {Nature}\ }\textbf {\bibinfo {volume} {406}},\ \bibinfo
  {pages} {299--302} (\bibinfo {year} {2000})}\BibitemShut {NoStop}%
\bibitem [{\citenamefont {Eder}\ \emph {et~al.}(2012)\citenamefont {Eder},
  \citenamefont {Cadiou}, \citenamefont {Muhamad}, \citenamefont {McNaughton},
  \citenamefont {Kirschvink},\ and\ \citenamefont
  {Winklhofer}}]{eder2012magnetic}%
  \BibitemOpen
  \bibfield  {author} {\bibinfo {author} {\bibfnamefont {S.~H.}\ \bibnamefont
  {Eder}}, \bibinfo {author} {\bibfnamefont {H.}~\bibnamefont {Cadiou}},
  \bibinfo {author} {\bibfnamefont {A.}~\bibnamefont {Muhamad}}, \bibinfo
  {author} {\bibfnamefont {P.~A.}\ \bibnamefont {McNaughton}}, \bibinfo
  {author} {\bibfnamefont {J.~L.}\ \bibnamefont {Kirschvink}}, \ and\ \bibinfo
  {author} {\bibfnamefont {M.}~\bibnamefont {Winklhofer}},\ }\bibfield  {title}
  {\enquote {\bibinfo {title} {Magnetic characterization of isolated candidate
  vertebrate magnetoreceptor cells},}\ }\href {\doibase
  10.1073/pnas.1205653109} {\bibfield  {journal} {\bibinfo  {journal} {Proc.
  Natl. Acad. Sci. U.S.A.}\ }\textbf {\bibinfo {volume} {109}},\ \bibinfo
  {pages} {12022--12027} (\bibinfo {year} {2012})}\BibitemShut {NoStop}%
\bibitem [{\citenamefont {Edelman}\ \emph {et~al.}(2015)\citenamefont
  {Edelman}, \citenamefont {Fritz}, \citenamefont {Nimpf}, \citenamefont
  {Pichler}, \citenamefont {Lauwers}, \citenamefont {Hickman}, \citenamefont
  {Papadaki-Anastasopoulou}, \citenamefont {Ushakova}, \citenamefont {Heuser},
  \citenamefont {Resch} \emph {et~al.}}]{edelman2015no}%
  \BibitemOpen
  \bibfield  {author} {\bibinfo {author} {\bibfnamefont {N.~B.}\ \bibnamefont
  {Edelman}}, \bibinfo {author} {\bibfnamefont {T.}~\bibnamefont {Fritz}},
  \bibinfo {author} {\bibfnamefont {S.}~\bibnamefont {Nimpf}}, \bibinfo
  {author} {\bibfnamefont {P.}~\bibnamefont {Pichler}}, \bibinfo {author}
  {\bibfnamefont {M.}~\bibnamefont {Lauwers}}, \bibinfo {author} {\bibfnamefont
  {R.~W.}\ \bibnamefont {Hickman}}, \bibinfo {author} {\bibfnamefont
  {A.}~\bibnamefont {Papadaki-Anastasopoulou}}, \bibinfo {author}
  {\bibfnamefont {L.}~\bibnamefont {Ushakova}}, \bibinfo {author}
  {\bibfnamefont {T.}~\bibnamefont {Heuser}}, \bibinfo {author} {\bibfnamefont
  {G.~P.}\ \bibnamefont {Resch}},  \emph {et~al.},\ }\bibfield  {title}
  {\enquote {\bibinfo {title} {No evidence for intracellular magnetite in
  putative vertebrate magnetoreceptors identified by magnetic screening},}\
  }\href {\doibase 10.1073/pnas.1407915112} {\bibfield  {journal} {\bibinfo
  {journal} {Proc. Natl. Acad. Sci. U.S.A.}\ }\textbf {\bibinfo {volume}
  {112}},\ \bibinfo {pages} {262--267} (\bibinfo {year} {2015})}\BibitemShut
  {NoStop}%
\bibitem [{\citenamefont {Doherty}\ \emph {et~al.}(2013)\citenamefont
  {Doherty}, \citenamefont {Manson}, \citenamefont {Delaney}, \citenamefont
  {Jelezko}, \citenamefont {Wrachtrup},\ and\ \citenamefont
  {Hollenberg}}]{Doherty2013}%
  \BibitemOpen
  \bibfield  {author} {\bibinfo {author} {\bibfnamefont {M.~W.}\ \bibnamefont
  {Doherty}}, \bibinfo {author} {\bibfnamefont {N.~B.}\ \bibnamefont {Manson}},
  \bibinfo {author} {\bibfnamefont {P.}~\bibnamefont {Delaney}}, \bibinfo
  {author} {\bibfnamefont {F.}~\bibnamefont {Jelezko}}, \bibinfo {author}
  {\bibfnamefont {J.}~\bibnamefont {Wrachtrup}}, \ and\ \bibinfo {author}
  {\bibfnamefont {L.~C.}\ \bibnamefont {Hollenberg}},\ }\bibfield  {title}
  {\enquote {\bibinfo {title} {The nitrogen-vacancy colour centre in
  diamond},}\ }\href {\doibase https://doi.org/10.1016/j.physrep.2013.02.001}
  {\bibfield  {journal} {\bibinfo  {journal} {Phys. Rep.}\ }\textbf {\bibinfo
  {volume} {528}},\ \bibinfo {pages} {1--45} (\bibinfo {year}
  {2013})}\BibitemShut {NoStop}%
\bibitem [{\citenamefont {Kehayias}\ \emph {et~al.}(2019)\citenamefont
  {Kehayias}, \citenamefont {Turner}, \citenamefont {Trubko}, \citenamefont
  {Schloss}, \citenamefont {Hart}, \citenamefont {Wesson}, \citenamefont
  {Glenn},\ and\ \citenamefont {Walsworth}}]{Kehayias2019stress}%
  \BibitemOpen
  \bibfield  {author} {\bibinfo {author} {\bibfnamefont {P.}~\bibnamefont
  {Kehayias}}, \bibinfo {author} {\bibfnamefont {M.~J.}\ \bibnamefont
  {Turner}}, \bibinfo {author} {\bibfnamefont {R.}~\bibnamefont {Trubko}},
  \bibinfo {author} {\bibfnamefont {J.~M.}\ \bibnamefont {Schloss}}, \bibinfo
  {author} {\bibfnamefont {C.~A.}\ \bibnamefont {Hart}}, \bibinfo {author}
  {\bibfnamefont {M.}~\bibnamefont {Wesson}}, \bibinfo {author} {\bibfnamefont
  {D.~R.}\ \bibnamefont {Glenn}}, \ and\ \bibinfo {author} {\bibfnamefont
  {R.~L.}\ \bibnamefont {Walsworth}},\ }\bibfield  {title} {\enquote {\bibinfo
  {title} {Imaging crystal stress in diamond using ensembles of
  nitrogen-vacancy centers},}\ }\href {\doibase 10.1103/PhysRevB.100.174103}
  {\bibfield  {journal} {\bibinfo  {journal} {Phys. Rev. B}\ }\textbf {\bibinfo
  {volume} {100}},\ \bibinfo {pages} {174103} (\bibinfo {year}
  {2019})}\BibitemShut {NoStop}%
\bibitem [{\citenamefont {Udvarhelyi}\ \emph {et~al.}(2018)\citenamefont
  {Udvarhelyi}, \citenamefont {Shkolnikov}, \citenamefont {Gali}, \citenamefont
  {Burkard},\ and\ \citenamefont {P\'alyi}}]{Udvarhelyi2018spinstrain}%
  \BibitemOpen
  \bibfield  {author} {\bibinfo {author} {\bibfnamefont {P.}~\bibnamefont
  {Udvarhelyi}}, \bibinfo {author} {\bibfnamefont {V.~O.}\ \bibnamefont
  {Shkolnikov}}, \bibinfo {author} {\bibfnamefont {A.}~\bibnamefont {Gali}},
  \bibinfo {author} {\bibfnamefont {G.}~\bibnamefont {Burkard}}, \ and\
  \bibinfo {author} {\bibfnamefont {A.}~\bibnamefont {P\'alyi}},\ }\bibfield
  {title} {\enquote {\bibinfo {title} {Spin-strain interaction in
  nitrogen-vacancy centers in diamond},}\ }\href {\doibase
  10.1103/PhysRevB.98.075201} {\bibfield  {journal} {\bibinfo  {journal} {Phys.
  Rev. B}\ }\textbf {\bibinfo {volume} {98}},\ \bibinfo {pages} {075201}
  (\bibinfo {year} {2018})}\BibitemShut {NoStop}%
\bibitem [{\citenamefont {Acosta}\ \emph {et~al.}(2010)\citenamefont {Acosta},
  \citenamefont {Bauch}, \citenamefont {Ledbetter}, \citenamefont {Waxman},
  \citenamefont {Bouchard},\ and\ \citenamefont {Budker}}]{Acosta2010}%
  \BibitemOpen
  \bibfield  {author} {\bibinfo {author} {\bibfnamefont {V.~M.}\ \bibnamefont
  {Acosta}}, \bibinfo {author} {\bibfnamefont {E.}~\bibnamefont {Bauch}},
  \bibinfo {author} {\bibfnamefont {M.~P.}\ \bibnamefont {Ledbetter}}, \bibinfo
  {author} {\bibfnamefont {A.}~\bibnamefont {Waxman}}, \bibinfo {author}
  {\bibfnamefont {L.-S.}\ \bibnamefont {Bouchard}}, \ and\ \bibinfo {author}
  {\bibfnamefont {D.}~\bibnamefont {Budker}},\ }\bibfield  {title} {\enquote
  {\bibinfo {title} {Temperature dependence of the nitrogen-vacancy magnetic
  resonance in diamond},}\ }\href {\doibase 10.1103/PhysRevLett.104.070801}
  {\bibfield  {journal} {\bibinfo  {journal} {Phys. Rev. Lett.}\ }\textbf
  {\bibinfo {volume} {104}},\ \bibinfo {pages} {070801} (\bibinfo {year}
  {2010})}\BibitemShut {NoStop}%
\bibitem [{\citenamefont {Mamin}\ \emph {et~al.}(2014)\citenamefont {Mamin},
  \citenamefont {Sherwood}, \citenamefont {Kim}, \citenamefont {Rettner},
  \citenamefont {Ohno}, \citenamefont {Awschalom},\ and\ \citenamefont
  {Rugar}}]{mamin2014multipulse}%
  \BibitemOpen
  \bibfield  {author} {\bibinfo {author} {\bibfnamefont {H.}~\bibnamefont
  {Mamin}}, \bibinfo {author} {\bibfnamefont {M.}~\bibnamefont {Sherwood}},
  \bibinfo {author} {\bibfnamefont {M.}~\bibnamefont {Kim}}, \bibinfo {author}
  {\bibfnamefont {C.}~\bibnamefont {Rettner}}, \bibinfo {author} {\bibfnamefont
  {K.}~\bibnamefont {Ohno}}, \bibinfo {author} {\bibfnamefont {D.}~\bibnamefont
  {Awschalom}}, \ and\ \bibinfo {author} {\bibfnamefont {D.}~\bibnamefont
  {Rugar}},\ }\bibfield  {title} {\enquote {\bibinfo {title} {Multipulse
  double-quantum magnetometry with near-surface nitrogen-vacancy centers},}\
  }\href {\doibase https://doi.org/10.1103/PhysRevLett.113.030803} {\bibfield
  {journal} {\bibinfo  {journal} {Phys. Rev. Lett.}\ }\textbf {\bibinfo
  {volume} {113}},\ \bibinfo {pages} {030803} (\bibinfo {year}
  {2014})}\BibitemShut {NoStop}%
\bibitem [{\citenamefont {Fu}\ \emph {et~al.}(2023)\citenamefont {Fu},
  \citenamefont {Maher}, \citenamefont {Nie}, \citenamefont {Gao},
  \citenamefont {Berndt}, \citenamefont {Folsom},\ and\ \citenamefont
  {Cavanaugh}}]{fu2023pinpointing}%
  \BibitemOpen
  \bibfield  {author} {\bibinfo {author} {\bibfnamefont {R.~R.}\ \bibnamefont
  {Fu}}, \bibinfo {author} {\bibfnamefont {B.~A.}\ \bibnamefont {Maher}},
  \bibinfo {author} {\bibfnamefont {J.}~\bibnamefont {Nie}}, \bibinfo {author}
  {\bibfnamefont {P.}~\bibnamefont {Gao}}, \bibinfo {author} {\bibfnamefont
  {T.}~\bibnamefont {Berndt}}, \bibinfo {author} {\bibfnamefont
  {E.}~\bibnamefont {Folsom}}, \ and\ \bibinfo {author} {\bibfnamefont
  {T.}~\bibnamefont {Cavanaugh}},\ }\bibfield  {title} {\enquote {\bibinfo
  {title} {Pinpointing the mechanism of magnetic enhancement in modern soils
  using high-resolution magnetic field imaging},}\ }\href {\doibase
  https://doi.org/10.1029/2022GC010812} {\bibfield  {journal} {\bibinfo
  {journal} {Geochem. Geophys. Geosyst.}\ }\textbf {\bibinfo {volume} {24}},\
  \bibinfo {pages} {e2022GC010812} (\bibinfo {year} {2023})}\BibitemShut
  {NoStop}%
\bibitem [{\citenamefont {Marshall}\ \emph {et~al.}(2022)\citenamefont
  {Marshall}, \citenamefont {Ebadi}, \citenamefont {Hart}, \citenamefont
  {Turner}, \citenamefont {Ku}, \citenamefont {Phillips},\ and\ \citenamefont
  {Walsworth}}]{Marshall2022precision}%
  \BibitemOpen
  \bibfield  {author} {\bibinfo {author} {\bibfnamefont {M.~C.}\ \bibnamefont
  {Marshall}}, \bibinfo {author} {\bibfnamefont {R.}~\bibnamefont {Ebadi}},
  \bibinfo {author} {\bibfnamefont {C.}~\bibnamefont {Hart}}, \bibinfo {author}
  {\bibfnamefont {M.~J.}\ \bibnamefont {Turner}}, \bibinfo {author}
  {\bibfnamefont {M.~J.}\ \bibnamefont {Ku}}, \bibinfo {author} {\bibfnamefont
  {D.~F.}\ \bibnamefont {Phillips}}, \ and\ \bibinfo {author} {\bibfnamefont
  {R.~L.}\ \bibnamefont {Walsworth}},\ }\bibfield  {title} {\enquote {\bibinfo
  {title} {High-precision mapping of diamond crystal strain using quantum
  interferometry},}\ }\href {\doibase 10.1103/PhysRevApplied.17.024041}
  {\bibfield  {journal} {\bibinfo  {journal} {Phys. Rev. Appl.}\ }\textbf
  {\bibinfo {volume} {17}},\ \bibinfo {pages} {024041} (\bibinfo {year}
  {2022})}\BibitemShut {NoStop}%
\bibitem [{com(2021)}]{comsol}%
  \BibitemOpen
  \href {http://www.comsol.com/products/multiphysics/} {\enquote {\bibinfo
  {title} {{COMSOL} {Multiphysics}\textsuperscript{\textregistered} v. 6.0},}\
  } (\bibinfo {year} {2021})\BibitemShut {NoStop}%
\bibitem [{\citenamefont {Kehayias}\ \emph {et~al.}(2022)\citenamefont
  {Kehayias}, \citenamefont {Levine}, \citenamefont {Basso}, \citenamefont
  {Henshaw}, \citenamefont {Ziabari}, \citenamefont {Titze}, \citenamefont
  {Haltli}, \citenamefont {Okoro}, \citenamefont {Tibbetts}, \citenamefont
  {Udoni} \emph {et~al.}}]{kehayias2022measurement}%
  \BibitemOpen
  \bibfield  {author} {\bibinfo {author} {\bibfnamefont {P.}~\bibnamefont
  {Kehayias}}, \bibinfo {author} {\bibfnamefont {E.}~\bibnamefont {Levine}},
  \bibinfo {author} {\bibfnamefont {L.}~\bibnamefont {Basso}}, \bibinfo
  {author} {\bibfnamefont {J.}~\bibnamefont {Henshaw}}, \bibinfo {author}
  {\bibfnamefont {M.~S.}\ \bibnamefont {Ziabari}}, \bibinfo {author}
  {\bibfnamefont {M.}~\bibnamefont {Titze}}, \bibinfo {author} {\bibfnamefont
  {R.}~\bibnamefont {Haltli}}, \bibinfo {author} {\bibfnamefont
  {J.}~\bibnamefont {Okoro}}, \bibinfo {author} {\bibfnamefont
  {D.}~\bibnamefont {Tibbetts}}, \bibinfo {author} {\bibfnamefont
  {D.}~\bibnamefont {Udoni}},  \emph {et~al.},\ }\bibfield  {title} {\enquote
  {\bibinfo {title} {Measurement and simulation of the magnetic fields from a
  555 timer integrated circuit using a quantum diamond microscope and
  finite-element analysis},}\ }\href@noop {} {\bibfield  {journal} {\bibinfo
  {journal} {Phys. Rev. Appl.}\ }\textbf {\bibinfo {volume} {17}},\ \bibinfo
  {pages} {014021} (\bibinfo {year} {2022})}\BibitemShut {NoStop}%
\bibitem [{\citenamefont {Buades}, \citenamefont {Coll},\ and\ \citenamefont
  {Morel}(2005{\natexlab{a}})}]{buades2005non}%
  \BibitemOpen
  \bibfield  {author} {\bibinfo {author} {\bibfnamefont {A.}~\bibnamefont
  {Buades}}, \bibinfo {author} {\bibfnamefont {B.}~\bibnamefont {Coll}}, \ and\
  \bibinfo {author} {\bibfnamefont {J.-M.}\ \bibnamefont {Morel}},\ }\bibfield
  {title} {\enquote {\bibinfo {title} {A non-local algorithm for image
  denoising},}\ }in\ \href {\doibase 10.1109/CVPR.2005.38} {\emph {\bibinfo
  {booktitle} {2005 IEEE Comput. Soc. Conf. Comput. Vis. Pattern Recognit.
  (CVPR)}}},\ Vol.~\bibinfo {volume} {2}\ (\bibinfo {organization} {IEEE},\
  \bibinfo {year} {2005})\ pp.\ \bibinfo {pages} {60--65}\BibitemShut {NoStop}%
\bibitem [{\citenamefont {Buades}, \citenamefont {Coll},\ and\ \citenamefont
  {Morel}(2005{\natexlab{b}})}]{buades2005review}%
  \BibitemOpen
  \bibfield  {author} {\bibinfo {author} {\bibfnamefont {A.}~\bibnamefont
  {Buades}}, \bibinfo {author} {\bibfnamefont {B.}~\bibnamefont {Coll}}, \ and\
  \bibinfo {author} {\bibfnamefont {J.-M.}\ \bibnamefont {Morel}},\ }\bibfield
  {title} {\enquote {\bibinfo {title} {A review of image denoising algorithms,
  with a new one},}\ }\href {\doibase https://doi.org/10.1137/040616024}
  {\bibfield  {journal} {\bibinfo  {journal} {Multiscale Model. Simul.}\
  }\textbf {\bibinfo {volume} {4}},\ \bibinfo {pages} {490--530} (\bibinfo
  {year} {2005}{\natexlab{b}})}\BibitemShut {NoStop}%
\bibitem [{\citenamefont {Bradski}(2000)}]{opencv_library}%
  \BibitemOpen
  \bibfield  {author} {\bibinfo {author} {\bibfnamefont {G.}~\bibnamefont
  {Bradski}},\ }\bibfield  {title} {\enquote {\bibinfo {title} {{The OpenCV
  Library}},}\ }\href@noop {} {\bibfield  {journal} {\bibinfo  {journal} {Dr.
  Dobb's Journal of Software Tools}\ } (\bibinfo {year} {2000})}\BibitemShut
  {NoStop}%
\bibitem [{\citenamefont {Winklhofer}\ and\ \citenamefont
  {Kirschvink}(2010)}]{winklhofer2010quantitative}%
  \BibitemOpen
  \bibfield  {author} {\bibinfo {author} {\bibfnamefont {M.}~\bibnamefont
  {Winklhofer}}\ and\ \bibinfo {author} {\bibfnamefont {J.~L.}\ \bibnamefont
  {Kirschvink}},\ }\bibfield  {title} {\enquote {\bibinfo {title} {A
  quantitative assessment of torque-transducer models for magnetoreception},}\
  }\href {\doibase 10.1098/rsif.2009.0435.focus} {\bibfield  {journal}
  {\bibinfo  {journal} {J. R. Soc. Interface}\ }\textbf {\bibinfo {volume}
  {7}},\ \bibinfo {pages} {S273--S289} (\bibinfo {year} {2010})}\BibitemShut
  {NoStop}%
\bibitem [{\citenamefont {Walker}\ \emph {et~al.}(1997)\citenamefont {Walker},
  \citenamefont {Diebel}, \citenamefont {Haugh}, \citenamefont {Pankhurst},
  \citenamefont {Montgomery},\ and\ \citenamefont
  {Green}}]{walker1997structure}%
  \BibitemOpen
  \bibfield  {author} {\bibinfo {author} {\bibfnamefont {M.~M.}\ \bibnamefont
  {Walker}}, \bibinfo {author} {\bibfnamefont {C.~E.}\ \bibnamefont {Diebel}},
  \bibinfo {author} {\bibfnamefont {C.~V.}\ \bibnamefont {Haugh}}, \bibinfo
  {author} {\bibfnamefont {P.~M.}\ \bibnamefont {Pankhurst}}, \bibinfo {author}
  {\bibfnamefont {J.~C.}\ \bibnamefont {Montgomery}}, \ and\ \bibinfo {author}
  {\bibfnamefont {C.~R.}\ \bibnamefont {Green}},\ }\bibfield  {title} {\enquote
  {\bibinfo {title} {Structure and function of the vertebrate magnetic
  sense},}\ }\href {\doibase https://www.nature.com/articles/37057} {\bibfield
  {journal} {\bibinfo  {journal} {Nature}\ }\textbf {\bibinfo {volume} {390}},\
  \bibinfo {pages} {371--376} (\bibinfo {year} {1997})}\BibitemShut {NoStop}%
\bibitem [{\citenamefont {J{\ae}ger}\ and\ \citenamefont
  {Tveito}(2022)}]{jaeger2022deriving}%
  \BibitemOpen
  \bibfield  {author} {\bibinfo {author} {\bibfnamefont {K.~H.}\ \bibnamefont
  {J{\ae}ger}}\ and\ \bibinfo {author} {\bibfnamefont {A.}~\bibnamefont
  {Tveito}},\ }\bibfield  {title} {\enquote {\bibinfo {title} {Deriving the
  bidomain model of cardiac electrophysiology from a cell-based model;
  properties and comparisons},}\ }\href {\doibase
  https://www.frontiersin.org/articles/10.3389/fphys.2021.811029/full}
  {\bibfield  {journal} {\bibinfo  {journal} {Front. Physiol.}\ }\textbf
  {\bibinfo {volume} {12}},\ \bibinfo {pages} {2439} (\bibinfo {year}
  {2022})}\BibitemShut {NoStop}%
\bibitem [{\citenamefont {Rohr}(2004)}]{rohr2004role}%
  \BibitemOpen
  \bibfield  {author} {\bibinfo {author} {\bibfnamefont {S.}~\bibnamefont
  {Rohr}},\ }\bibfield  {title} {\enquote {\bibinfo {title} {Role of gap
  junctions in the propagation of the cardiac action potential},}\ }\href
  {\doibase https://academic.oup.com/cardiovascres/article/62/2/309/316416}
  {\bibfield  {journal} {\bibinfo  {journal} {Cardiovasc. Res.}\ }\textbf
  {\bibinfo {volume} {62}},\ \bibinfo {pages} {309--322} (\bibinfo {year}
  {2004})}\BibitemShut {NoStop}%
\bibitem [{\citenamefont {Dhein}\ and\ \citenamefont
  {Salameh}(2021)}]{dhein2021remodeling}%
  \BibitemOpen
  \bibfield  {author} {\bibinfo {author} {\bibfnamefont {S.}~\bibnamefont
  {Dhein}}\ and\ \bibinfo {author} {\bibfnamefont {A.}~\bibnamefont
  {Salameh}},\ }\bibfield  {title} {\enquote {\bibinfo {title} {Remodeling of
  cardiac gap junctional cell--cell coupling},}\ }\href {\doibase
  10.3390/cells10092422} {\bibfield  {journal} {\bibinfo  {journal} {Cells}\
  }\textbf {\bibinfo {volume} {10}},\ \bibinfo {pages} {2422} (\bibinfo {year}
  {2021})}\BibitemShut {NoStop}%
\bibitem [{\citenamefont {Barral}\ and\ \citenamefont
  {Kurian}(2016)}]{barral2016utility}%
  \BibitemOpen
  \bibfield  {author} {\bibinfo {author} {\bibfnamefont {S.}~\bibnamefont
  {Barral}}\ and\ \bibinfo {author} {\bibfnamefont {M.~A.}\ \bibnamefont
  {Kurian}},\ }\bibfield  {title} {\enquote {\bibinfo {title} {Utility of
  induced pluripotent stem cells for the study and treatment of genetic
  diseases: focus on childhood neurological disorders},}\ }\href {\doibase
  10.3389/fnmol.2016.00078} {\bibfield  {journal} {\bibinfo  {journal} {Front.
  Mol. Neurosci.}\ }\textbf {\bibinfo {volume} {9}},\ \bibinfo {pages} {78}
  (\bibinfo {year} {2016})}\BibitemShut {NoStop}%
\bibitem [{\citenamefont {Holzer}\ \emph {et~al.}(2004)\citenamefont {Holzer},
  \citenamefont {Fong}, \citenamefont {Sidorov}, \citenamefont {Wikswo},\ and\
  \citenamefont {Baudenbacher}}]{Holzer2004-magneticimagecardiac}%
  \BibitemOpen
  \bibfield  {author} {\bibinfo {author} {\bibfnamefont {J.~R.}\ \bibnamefont
  {Holzer}}, \bibinfo {author} {\bibfnamefont {L.~E.}\ \bibnamefont {Fong}},
  \bibinfo {author} {\bibfnamefont {V.~Y.}\ \bibnamefont {Sidorov}}, \bibinfo
  {author} {\bibfnamefont {J.~P.}\ \bibnamefont {Wikswo}}, \ and\ \bibinfo
  {author} {\bibfnamefont {F.}~\bibnamefont {Baudenbacher}},\ }\bibfield
  {title} {\enquote {\bibinfo {title} {High resolution magnetic images of
  planar wave fronts reveal bidomain properties of cardiac tissue},}\ }\href
  {\doibase https://doi.org/10.1529/biophysj.104.049163} {\bibfield  {journal}
  {\bibinfo  {journal} {Biophys. J.}\ }\textbf {\bibinfo {volume} {87}},\
  \bibinfo {pages} {4326--4332} (\bibinfo {year} {2004})}\BibitemShut {NoStop}%
\bibitem [{\citenamefont {Rossi}\ \emph {et~al.}(2017)\citenamefont {Rossi},
  \citenamefont {Buccarello}, \citenamefont {Ershler}, \citenamefont {Lux},
  \citenamefont {Callegari}, \citenamefont {Corradi}, \citenamefont
  {Carnevali}, \citenamefont {Sgoifo}, \citenamefont {Miragoli}, \citenamefont
  {Musso} \emph {et~al.}}]{rossi2017effect}%
  \BibitemOpen
  \bibfield  {author} {\bibinfo {author} {\bibfnamefont {S.}~\bibnamefont
  {Rossi}}, \bibinfo {author} {\bibfnamefont {A.}~\bibnamefont {Buccarello}},
  \bibinfo {author} {\bibfnamefont {P.~R.}\ \bibnamefont {Ershler}}, \bibinfo
  {author} {\bibfnamefont {R.~L.}\ \bibnamefont {Lux}}, \bibinfo {author}
  {\bibfnamefont {S.}~\bibnamefont {Callegari}}, \bibinfo {author}
  {\bibfnamefont {D.}~\bibnamefont {Corradi}}, \bibinfo {author} {\bibfnamefont
  {L.}~\bibnamefont {Carnevali}}, \bibinfo {author} {\bibfnamefont
  {A.}~\bibnamefont {Sgoifo}}, \bibinfo {author} {\bibfnamefont
  {M.}~\bibnamefont {Miragoli}}, \bibinfo {author} {\bibfnamefont
  {E.}~\bibnamefont {Musso}},  \emph {et~al.},\ }\bibfield  {title} {\enquote
  {\bibinfo {title} {Effect of anisotropy on ventricular vulnerability to
  unidirectional block and reentry by single premature stimulation during
  normal sinus rhythm in rat heart},}\ }\href {\doibase
  https://doi.org/10.1152/ajpheart.00366.2016} {\bibfield  {journal} {\bibinfo
  {journal} {Am. J. Physiol. Heart Circ. Physiol.}\ }\textbf {\bibinfo {volume}
  {312}},\ \bibinfo {pages} {H584--H607} (\bibinfo {year} {2017})}\BibitemShut
  {NoStop}%
\bibitem [{\citenamefont {Sepulveda}, \citenamefont {Roth},\ and\ \citenamefont
  {Wikswo}(1989)}]{sepulveda1989current}%
  \BibitemOpen
  \bibfield  {author} {\bibinfo {author} {\bibfnamefont {N.~G.}\ \bibnamefont
  {Sepulveda}}, \bibinfo {author} {\bibfnamefont {B.~J.}\ \bibnamefont {Roth}},
  \ and\ \bibinfo {author} {\bibfnamefont {J.}~\bibnamefont {Wikswo}},\
  }\bibfield  {title} {\enquote {\bibinfo {title} {Current injection into a
  two-dimensional anisotropic bidomain},}\ }\href {\doibase
  https://doi.org/10.1016/S0006-3495(89)82897-8} {\bibfield  {journal}
  {\bibinfo  {journal} {Biophys. J.}\ }\textbf {\bibinfo {volume} {55}},\
  \bibinfo {pages} {987--999} (\bibinfo {year} {1989})}\BibitemShut {NoStop}%
\bibitem [{\citenamefont {Rossi}\ and\ \citenamefont
  {Griffith}(2017)}]{rossi2017incorporating}%
  \BibitemOpen
  \bibfield  {author} {\bibinfo {author} {\bibfnamefont {S.}~\bibnamefont
  {Rossi}}\ and\ \bibinfo {author} {\bibfnamefont {B.~E.}\ \bibnamefont
  {Griffith}},\ }\bibfield  {title} {\enquote {\bibinfo {title} {Incorporating
  inductances in tissue-scale models of cardiac electrophysiology},}\ }\href
  {\doibase https://doi.org/10.1063/1.5000706} {\bibfield  {journal} {\bibinfo
  {journal} {Chaos}\ }\textbf {\bibinfo {volume} {27}} (\bibinfo {year}
  {2017}),\ https://doi.org/10.1063/1.5000706}\BibitemShut {NoStop}%
\bibitem [{\citenamefont {Hochbaum}\ \emph {et~al.}(2014)\citenamefont
  {Hochbaum}, \citenamefont {Zhao}, \citenamefont {Farhi}, \citenamefont
  {Klapoetke}, \citenamefont {Werley}, \citenamefont {Kapoor}, \citenamefont
  {Zou}, \citenamefont {Kralj}, \citenamefont {Maclaurin}, \citenamefont
  {Smedemark-Margulies},\ and\ \citenamefont
  {et~al.}}]{Hochbaum2014-rhodopsins}%
  \BibitemOpen
  \bibfield  {author} {\bibinfo {author} {\bibfnamefont {D.~R.}\ \bibnamefont
  {Hochbaum}}, \bibinfo {author} {\bibfnamefont {Y.}~\bibnamefont {Zhao}},
  \bibinfo {author} {\bibfnamefont {S.~L.}\ \bibnamefont {Farhi}}, \bibinfo
  {author} {\bibfnamefont {N.}~\bibnamefont {Klapoetke}}, \bibinfo {author}
  {\bibfnamefont {C.~A.}\ \bibnamefont {Werley}}, \bibinfo {author}
  {\bibfnamefont {V.}~\bibnamefont {Kapoor}}, \bibinfo {author} {\bibfnamefont
  {P.}~\bibnamefont {Zou}}, \bibinfo {author} {\bibfnamefont {J.~M.}\
  \bibnamefont {Kralj}}, \bibinfo {author} {\bibfnamefont {D.}~\bibnamefont
  {Maclaurin}}, \bibinfo {author} {\bibfnamefont {N.}~\bibnamefont
  {Smedemark-Margulies}}, \ and\ \bibinfo {author} {\bibnamefont {et~al.}},\
  }\bibfield  {title} {\enquote {\bibinfo {title} {All-optical
  electrophysiology in mammalian neurons using engineered microbial
  rhodopsins},}\ }\href {\doibase 10.1038/nmeth.3000} {\bibfield  {journal}
  {\bibinfo  {journal} {Nat. Methods}\ }\textbf {\bibinfo {volume} {11}},\
  \bibinfo {pages} {825–833} (\bibinfo {year} {2014})}\BibitemShut {NoStop}%
\bibitem [{\citenamefont {Sundaram}\ \emph {et~al.}(2016)\citenamefont
  {Sundaram}, \citenamefont {Nummenmaa}, \citenamefont {Wells}, \citenamefont
  {Orbach}, \citenamefont {Orringer}, \citenamefont {Mulkern},\ and\
  \citenamefont {Okada}}]{sundaram2016direct}%
  \BibitemOpen
  \bibfield  {author} {\bibinfo {author} {\bibfnamefont {P.}~\bibnamefont
  {Sundaram}}, \bibinfo {author} {\bibfnamefont {A.}~\bibnamefont {Nummenmaa}},
  \bibinfo {author} {\bibfnamefont {W.}~\bibnamefont {Wells}}, \bibinfo
  {author} {\bibfnamefont {D.}~\bibnamefont {Orbach}}, \bibinfo {author}
  {\bibfnamefont {D.}~\bibnamefont {Orringer}}, \bibinfo {author}
  {\bibfnamefont {R.}~\bibnamefont {Mulkern}}, \ and\ \bibinfo {author}
  {\bibfnamefont {Y.}~\bibnamefont {Okada}},\ }\bibfield  {title} {\enquote
  {\bibinfo {title} {Direct neural current imaging in an intact cerebellum with
  magnetic resonance imaging},}\ }\href {\doibase
  https://doi.org/10.1016/j.neuroimage.2016.01.059} {\bibfield  {journal}
  {\bibinfo  {journal} {Neuroimage}\ }\textbf {\bibinfo {volume} {132}},\
  \bibinfo {pages} {477--490} (\bibinfo {year} {2016})}\BibitemShut {NoStop}%
\bibitem [{\citenamefont {Ihle}\ \emph {et~al.}(2022)\citenamefont {Ihle},
  \citenamefont {Girardin}, \citenamefont {Felder}, \citenamefont {Ruff},
  \citenamefont {Hengsteler}, \citenamefont {Duru}, \citenamefont {Weaver},
  \citenamefont {Forr{\'o}},\ and\ \citenamefont
  {V{\"o}r{\"o}s}}]{ihle2022experimental}%
  \BibitemOpen
  \bibfield  {author} {\bibinfo {author} {\bibfnamefont {S.~J.}\ \bibnamefont
  {Ihle}}, \bibinfo {author} {\bibfnamefont {S.}~\bibnamefont {Girardin}},
  \bibinfo {author} {\bibfnamefont {T.}~\bibnamefont {Felder}}, \bibinfo
  {author} {\bibfnamefont {T.}~\bibnamefont {Ruff}}, \bibinfo {author}
  {\bibfnamefont {J.}~\bibnamefont {Hengsteler}}, \bibinfo {author}
  {\bibfnamefont {J.}~\bibnamefont {Duru}}, \bibinfo {author} {\bibfnamefont
  {S.}~\bibnamefont {Weaver}}, \bibinfo {author} {\bibfnamefont
  {C.}~\bibnamefont {Forr{\'o}}}, \ and\ \bibinfo {author} {\bibfnamefont
  {J.}~\bibnamefont {V{\"o}r{\"o}s}},\ }\bibfield  {title} {\enquote {\bibinfo
  {title} {An experimental paradigm to investigate stimulation dependent
  activity in topologically constrained neuronal networks},}\ }\href {\doibase
  10.1016/j.bios.2021.113896} {\bibfield  {journal} {\bibinfo  {journal}
  {Biosens. Bioelectron.}\ }\textbf {\bibinfo {volume} {201}},\ \bibinfo
  {pages} {113896} (\bibinfo {year} {2022})}\BibitemShut {NoStop}%
\bibitem [{\citenamefont {Edmonds}\ \emph {et~al.}(2021)\citenamefont
  {Edmonds}, \citenamefont {Hart}, \citenamefont {Turner}, \citenamefont
  {Colard}, \citenamefont {Schloss}, \citenamefont {Olsson}, \citenamefont
  {Trubko}, \citenamefont {Markham}, \citenamefont {Rathmill}, \citenamefont
  {Horne-Smith} \emph {et~al.}}]{edmonds2021characterisation}%
  \BibitemOpen
  \bibfield  {author} {\bibinfo {author} {\bibfnamefont {A.~M.}\ \bibnamefont
  {Edmonds}}, \bibinfo {author} {\bibfnamefont {C.~A.}\ \bibnamefont {Hart}},
  \bibinfo {author} {\bibfnamefont {M.~J.}\ \bibnamefont {Turner}}, \bibinfo
  {author} {\bibfnamefont {P.-O.}\ \bibnamefont {Colard}}, \bibinfo {author}
  {\bibfnamefont {J.~M.}\ \bibnamefont {Schloss}}, \bibinfo {author}
  {\bibfnamefont {K.~S.}\ \bibnamefont {Olsson}}, \bibinfo {author}
  {\bibfnamefont {R.}~\bibnamefont {Trubko}}, \bibinfo {author} {\bibfnamefont
  {M.~L.}\ \bibnamefont {Markham}}, \bibinfo {author} {\bibfnamefont
  {A.}~\bibnamefont {Rathmill}}, \bibinfo {author} {\bibfnamefont
  {B.}~\bibnamefont {Horne-Smith}},  \emph {et~al.},\ }\bibfield  {title}
  {\enquote {\bibinfo {title} {Characterisation of {CVD} diamond with high
  concentrations of nitrogen for magnetic-field sensing applications},}\ }\href
  {\doibase 10.1088/2633-4356/abd88a} {\bibfield  {journal} {\bibinfo
  {journal} {Mater. Quantum. Technol.}\ }\textbf {\bibinfo {volume} {1}},\
  \bibinfo {pages} {025001} (\bibinfo {year} {2021})}\BibitemShut {NoStop}%
\bibitem [{\citenamefont {Wee}\ \emph {et~al.}(2007)\citenamefont {Wee},
  \citenamefont {Tzeng}, \citenamefont {Han}, \citenamefont {Chang},
  \citenamefont {Fann}, \citenamefont {Hsu}, \citenamefont {Chen},\ and\
  \citenamefont {Yu}}]{wee2007two}%
  \BibitemOpen
  \bibfield  {author} {\bibinfo {author} {\bibfnamefont {T.-L.}\ \bibnamefont
  {Wee}}, \bibinfo {author} {\bibfnamefont {Y.-K.}\ \bibnamefont {Tzeng}},
  \bibinfo {author} {\bibfnamefont {C.-C.}\ \bibnamefont {Han}}, \bibinfo
  {author} {\bibfnamefont {H.-C.}\ \bibnamefont {Chang}}, \bibinfo {author}
  {\bibfnamefont {W.}~\bibnamefont {Fann}}, \bibinfo {author} {\bibfnamefont
  {J.-H.}\ \bibnamefont {Hsu}}, \bibinfo {author} {\bibfnamefont {K.-M.}\
  \bibnamefont {Chen}}, \ and\ \bibinfo {author} {\bibfnamefont {Y.-C.}\
  \bibnamefont {Yu}},\ }\bibfield  {title} {\enquote {\bibinfo {title}
  {Two-photon excited fluorescence of nitrogen-vacancy centers in
  proton-irradiated type {I}b diamond},}\ }\href {\doibase
  https://doi.org/10.1021/jp073938o} {\bibfield  {journal} {\bibinfo  {journal}
  {J. Phys. Chem. A}\ }\textbf {\bibinfo {volume} {111}},\ \bibinfo {pages}
  {9379--9386} (\bibinfo {year} {2007})}\BibitemShut {NoStop}%
\bibitem [{\citenamefont {Balasubramanian}\ \emph {et~al.}(2019)\citenamefont
  {Balasubramanian}, \citenamefont {Osterkamp}, \citenamefont {Chen},
  \citenamefont {Chen}, \citenamefont {Teraji}, \citenamefont {Wu},
  \citenamefont {Naydenov},\ and\ \citenamefont
  {Jelezko}}]{balasubramanian2019dc}%
  \BibitemOpen
  \bibfield  {author} {\bibinfo {author} {\bibfnamefont {P.}~\bibnamefont
  {Balasubramanian}}, \bibinfo {author} {\bibfnamefont {C.}~\bibnamefont
  {Osterkamp}}, \bibinfo {author} {\bibfnamefont {Y.}~\bibnamefont {Chen}},
  \bibinfo {author} {\bibfnamefont {X.}~\bibnamefont {Chen}}, \bibinfo {author}
  {\bibfnamefont {T.}~\bibnamefont {Teraji}}, \bibinfo {author} {\bibfnamefont
  {E.}~\bibnamefont {Wu}}, \bibinfo {author} {\bibfnamefont {B.}~\bibnamefont
  {Naydenov}}, \ and\ \bibinfo {author} {\bibfnamefont {F.}~\bibnamefont
  {Jelezko}},\ }\bibfield  {title} {\enquote {\bibinfo {title} {dc magnetometry
  with engineered nitrogen-vacancy spin ensembles in diamond},}\ }\href
  {\doibase 10.1021/acs.nanolett.9b02993} {\bibfield  {journal} {\bibinfo
  {journal} {Nano Lett.}\ }\textbf {\bibinfo {volume} {19}},\ \bibinfo {pages}
  {6681--6686} (\bibinfo {year} {2019})}\BibitemShut {NoStop}%
\bibitem [{\citenamefont {DeVience}\ \emph {et~al.}(2015)\citenamefont
  {DeVience}, \citenamefont {Pham}, \citenamefont {Lovchinsky}, \citenamefont
  {Sushkov}, \citenamefont {Bar-Gill}, \citenamefont {Belthangady},
  \citenamefont {Casola}, \citenamefont {Corbett}, \citenamefont {Zhang},
  \citenamefont {Lukin} \emph {et~al.}}]{devience2015nanoscale}%
  \BibitemOpen
  \bibfield  {author} {\bibinfo {author} {\bibfnamefont {S.~J.}\ \bibnamefont
  {DeVience}}, \bibinfo {author} {\bibfnamefont {L.~M.}\ \bibnamefont {Pham}},
  \bibinfo {author} {\bibfnamefont {I.}~\bibnamefont {Lovchinsky}}, \bibinfo
  {author} {\bibfnamefont {A.~O.}\ \bibnamefont {Sushkov}}, \bibinfo {author}
  {\bibfnamefont {N.}~\bibnamefont {Bar-Gill}}, \bibinfo {author}
  {\bibfnamefont {C.}~\bibnamefont {Belthangady}}, \bibinfo {author}
  {\bibfnamefont {F.}~\bibnamefont {Casola}}, \bibinfo {author} {\bibfnamefont
  {M.}~\bibnamefont {Corbett}}, \bibinfo {author} {\bibfnamefont
  {H.}~\bibnamefont {Zhang}}, \bibinfo {author} {\bibfnamefont
  {M.}~\bibnamefont {Lukin}},  \emph {et~al.},\ }\bibfield  {title} {\enquote
  {\bibinfo {title} {Nanoscale {NMR} spectroscopy and imaging of multiple
  nuclear species},}\ }\href {\doibase 10.1038/nnano.2014.313} {\bibfield
  {journal} {\bibinfo  {journal} {Nature Nanotech.}\ }\textbf {\bibinfo
  {volume} {10}},\ \bibinfo {pages} {129--134} (\bibinfo {year}
  {2015})}\BibitemShut {NoStop}%
\bibitem [{\citenamefont {Arunkumar}\ \emph {et~al.}(2023)\citenamefont
  {Arunkumar}, \citenamefont {Olsson}, \citenamefont {Oon}, \citenamefont
  {Hart}, \citenamefont {Bucher}, \citenamefont {Glenn}, \citenamefont {Lukin},
  \citenamefont {Park}, \citenamefont {Ham},\ and\ \citenamefont
  {Walsworth}}]{arunkumar2023quantum}%
  \BibitemOpen
  \bibfield  {author} {\bibinfo {author} {\bibfnamefont {N.}~\bibnamefont
  {Arunkumar}}, \bibinfo {author} {\bibfnamefont {K.~S.}\ \bibnamefont
  {Olsson}}, \bibinfo {author} {\bibfnamefont {J.~T.}\ \bibnamefont {Oon}},
  \bibinfo {author} {\bibfnamefont {C.~A.}\ \bibnamefont {Hart}}, \bibinfo
  {author} {\bibfnamefont {D.~B.}\ \bibnamefont {Bucher}}, \bibinfo {author}
  {\bibfnamefont {D.~R.}\ \bibnamefont {Glenn}}, \bibinfo {author}
  {\bibfnamefont {M.~D.}\ \bibnamefont {Lukin}}, \bibinfo {author}
  {\bibfnamefont {H.}~\bibnamefont {Park}}, \bibinfo {author} {\bibfnamefont
  {D.}~\bibnamefont {Ham}}, \ and\ \bibinfo {author} {\bibfnamefont {R.~L.}\
  \bibnamefont {Walsworth}},\ }\bibfield  {title} {\enquote {\bibinfo {title}
  {Quantum logic enhanced sensing in solid-state spin ensembles},}\ }\href
  {\doibase https://doi.org/10.1103/PhysRevLett.131.100801} {\bibfield
  {journal} {\bibinfo  {journal} {Phys. Rev. Lett.}\ }\textbf {\bibinfo
  {volume} {131}},\ \bibinfo {pages} {100801} (\bibinfo {year}
  {2023})}\BibitemShut {NoStop}%
\bibitem [{\citenamefont {De~Lange}\ \emph {et~al.}(2012)\citenamefont
  {De~Lange}, \citenamefont {Van Der~Sar}, \citenamefont {Blok}, \citenamefont
  {Wang}, \citenamefont {Dobrovitski},\ and\ \citenamefont
  {Hanson}}]{de2012controlling}%
  \BibitemOpen
  \bibfield  {author} {\bibinfo {author} {\bibfnamefont {G.}~\bibnamefont
  {De~Lange}}, \bibinfo {author} {\bibfnamefont {T.}~\bibnamefont {Van
  Der~Sar}}, \bibinfo {author} {\bibfnamefont {M.}~\bibnamefont {Blok}},
  \bibinfo {author} {\bibfnamefont {Z.-H.}\ \bibnamefont {Wang}}, \bibinfo
  {author} {\bibfnamefont {V.}~\bibnamefont {Dobrovitski}}, \ and\ \bibinfo
  {author} {\bibfnamefont {R.}~\bibnamefont {Hanson}},\ }\bibfield  {title}
  {\enquote {\bibinfo {title} {Controlling the quantum dynamics of a mesoscopic
  spin bath in diamond},}\ }\href {\doibase 10.1038/srep00382} {\bibfield
  {journal} {\bibinfo  {journal} {Sci. Rep.}\ }\textbf {\bibinfo {volume}
  {2}},\ \bibinfo {pages} {1--5} (\bibinfo {year} {2012})}\BibitemShut
  {NoStop}%
\bibitem [{\citenamefont {Cox}, \citenamefont {Newton},\ and\ \citenamefont
  {Baker}(1994)}]{cox199413c}%
  \BibitemOpen
  \bibfield  {author} {\bibinfo {author} {\bibfnamefont {A.}~\bibnamefont
  {Cox}}, \bibinfo {author} {\bibfnamefont {M.}~\bibnamefont {Newton}}, \ and\
  \bibinfo {author} {\bibfnamefont {J.}~\bibnamefont {Baker}},\ }\bibfield
  {title} {\enquote {\bibinfo {title} {$^{13}${C}, $^{14}${N} and $^{15}${N}
  endor measurements on the single substitutional nitrogen centre ({P1}) in
  diamond},}\ }\href {\doibase 10.1088/0953-8984/6/2/025} {\bibfield  {journal}
  {\bibinfo  {journal} {Journal of Physics: Condensed Matter}\ }\textbf
  {\bibinfo {volume} {6}},\ \bibinfo {pages} {551} (\bibinfo {year}
  {1994})}\BibitemShut {NoStop}%
\bibitem [{hel(2023)}]{heliotis}%
  \BibitemOpen
  \href {https://www.heliotis.com/en/sensors/lock-in-camera/} {\enquote
  {\bibinfo {title} {{Heliotis AG} {heliCam}\textsuperscript{\texttrademark}
  {C3 Lock-in Camera}},}\ } (\bibinfo {year} {2023})\BibitemShut {NoStop}%
\end{thebibliography}%

\newpage
\onecolumngrid

\section*{Supplementary Material: Quantum Diamond Microscope for Dynamic Imaging of Magnetic Fields}
\setcounter{section}{0}
\setcounter{figure}{0}
\setcounter{equation}{0}
\renewcommand{\thesection}{S-\Roman{section}}
\renewcommand{\theequation}{S.\arabic{equation}}
\renewcommand{\thefigure}{S\arabic{figure}}

\section{Experimental Details} 
\label{supp:exp}

The QDM diamond sensor consists of a 10-\textmu m-thick, $^{15}$N-enriched CVD layer of NV centers ($[\text{N}]$~$=$~17\,ppm, $>$\,99.995\% $^{12}$C), grown by Element Six Ltd.~on a $2\times2\times0.5$\,mm$^3$ high-purity diamond substrate. Post-growth treatment via electron irradiation and annealing increases the NV concentration to about 2.4\,ppm.

 A diode-pumped solid-state laser (Lighthouse Photonics, Sprout-D5W) generates the
532\,nm laser beam. The laser beam is focused onto an acousto-optic modulator (AOM) (Gooch \& Housego, Model 3250-220) by a $f=500$\,mm spherical convex lens; and the first-order diffracted beam from the AOM is used for the experiments. To create optical pulses for NV initialization and readout, the radio frequency (RF) driver for the AOM is gated by switches (Mini-Circuits, ZASWA-2-50DR+) using transistor-transistor logic (TTL) pulses (SpinCore Technologies, PBESR-PRO-500). The first-order beam after the AOM is further shaped by an $f=130$\,mm cylindrical convex lens and a $f=100$\,mm spherical convex lens to produce an elliptical beam profile incident on a polished side surface of the diamond sample. Approximately 2.5\,W of laser power enters the diamond and undergoes total internal reflection at the surface of the NV layer, illuminating an area of about $300\times600$\,\textmu m$^2$. 
 NV fluorescence is collected by a 20$\times$/0.75\,NA Nikon objective, and filtered by a 647\,nm long-pass filter (Semrock) before being projected onto a photodiode (Thorlabs, PDA36A2) or a
 lock-in camera (Heliotis, heliCam C3). Photodiode voltage measurements are acquired using a DAQ system (National Instruments, NI USB6363). 
 
 Dual-tone microwave (MW) pulses for NV spin control are generated from two signal sources with built-in IQ mixers (Stanford Research System, SG384) and gated by switches and TTL pulses. A fabricated, nominally 300\,nm-thick, 800\,\textmu m-wide gold \textOmega-shaped coplanar waveguide delivers MW signals to the NV layer. The fabricated palladium wire phantom (in the shape of a terrapin) has a trace width $\approx$\,4\,\textmu m, spans an area of about $220\times180$\,\textmu m$^2$ area, and has a thickness of $\approx$\,10\,nm. To generate phantom magnetic-field signals, stimulus voltages produced by a waveform generator (Keysight, 33622A) are applied to the phantom to induce currents. 
 The phantom and MW loops are both fabricated on a silicon carbide wafer for mechanical support and heat dissipation. The diamond is glued to the surface of the MW loop; thus the minimal standoff between the phantom and NV surface is the thickness of the MW structure ($\approx$\,300\,nm). 
 
Multi-tone, resonant RF pulses for spin-bath driving are generated by an arbitrary waveform generator (Keysight, M8190) and gated by switches and TTL pulses. RF is delivered to the bath spins by a shorted co-axial loop ($\approx$\,1\,mm wide) placed above the MW waveguide.  
 
A bias magnetic field of about 4.3\,mT is provided by two pairs of temperature-compensated, permanent samarium cobalt ring magnets. This bias magnetic field magnitude is convenient with our present apparatus; but similar QDM performance should be possible for a wide range of magnetic fields, from millitesla to tesla scales. 

\begin{figure}
    \centering
    \includegraphics[width=8.5cm]{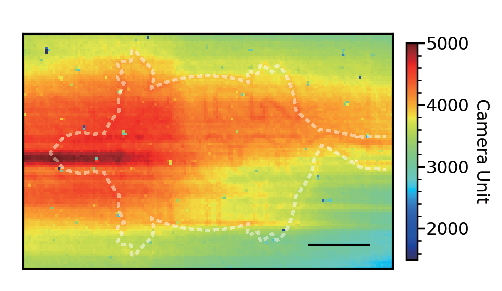}
    \caption{Optical image of the fabricated wire phantom. NV fluorescence intensity under continuous laser illumination is collected using the QDM's camera. Semi-transparent white dashed traces of the phantom perimeter are added as a guide to the eye. The measured fluorescence is reported in the camera unit, which is linearly proportional to the photon number. Scale bar: 50\,\textmu m.}
    \label{figS1}
\end{figure}

The size of a single camera pixel in the objective focal plane is determined using the known geometry of the phantom and its optical image as visualized using NV fluorescence.
The metallic surfaces of the phantom traces reflect the incident excitation laser light and amplify the local illumination intensity, resulting in enhanced NV fluorescence with a pattern following the phantom structure as shown in Figure~\ref{figS1}.
The NV layer area corresponding to each camera pixel is about $1.9\times1.9$\,\textmu m$^2$, which agrees with the estimate using the magnification of the objective (20$\times$) and the physical size of each pixel ($39.6\times39.6$\,\textmu m$^2$) as reported by the camera manufacturer.

The camera has a controllable internal exposure frequency of up to 1\,MHz and an external frame rate of up to 3.8\,kHz. Each external frame contains the net accumulated results over multiple internal exposures. Additional details regarding synchronizing camera internal exposure cycles with the dual-tone MW Ramsey sequence are provided in supplementary Section~\ref{supp:DQ4R}. The camera buffer can hold a continuous acquisition of 500 external frames. Data transfer from the camera buffer to a host computer takes about 5\,s. Additional software-related overhead, including data conversion and storage to integrate with other custom-made experimental control programs, consume about 2\,s for static magnetic-field imaging, which stores the average of 500 frames; and about 17s for dynamic magnetic-field imaging, which stores individual frames. This implementation-dependent overhead is not included in the acquisition time $T_{acq}$, but is accounted for in the wall-clock time $T_{wall}$.

\section{Ramsey Sequence for Magnetic Imaging} 
\label{supp:DQ4R}

The Ramsey sequence for magnetic imaging utilizes dual-tone MW pulses and a MW-phase alternation scheme to achieve uniform magnetic sensitivity over the field of view \cite{Hart2021-4Ramsey}.
The dual-tone MW pulse generates a double-quantum (DQ) coherence between the $\ket{m_s=+1}$ and $\ket{m_s=-1}$ ground state spin triplet sublevels of the NV center. 
Compared to using a single-quantum (SQ) coherence between $\ket{m_s=0}$ and $\ket{m_s=+1}$ or $\ket{m_s=-1}$, the DQ scheme is twice as sensitive to magnetic fields and can isolate magnetic signals from spurious strain and temperature artifacts. 
However, heterogeneity of MW control fields, diamond strain, and temperature can cause errors in DQ pulses across the NV ensemble and generate a residual SQ coherence.
Thus a single DQ Ramsey measurement is still not free from spurious signals unless pulse errors are mitigated. To overcome this challenge, the phases of the two MW tones can be alternated across four individual DQ Ramsey measurements (a protocol known as DQ 4-Ramsey \cite{Hart2021-4Ramsey}). If we denote the fluorescence signals from the $i$-th measurement as $S_i$, the phase-alternation scheme generates DQ magnetic signals  in $S_2$ and $S_4$ that are opposite to $S_1$ and $S_3$, while residual SQ signals in the sum $S_1+S_3$ are also observed in $S_2+S_4$. A final normalized signal $S_{norm}=S_1-S_2+S_3-S_4$ preserves the DQ component and achieves robustness to pulse errors and magnetic artifacts by eliminating the SQ contributions. In addition, the normalization scheme can be integrated with a lock-in camera to further mitigate laser intensity variation.

\begin{figure*}
    \centering
    \includegraphics[width=\linewidth]{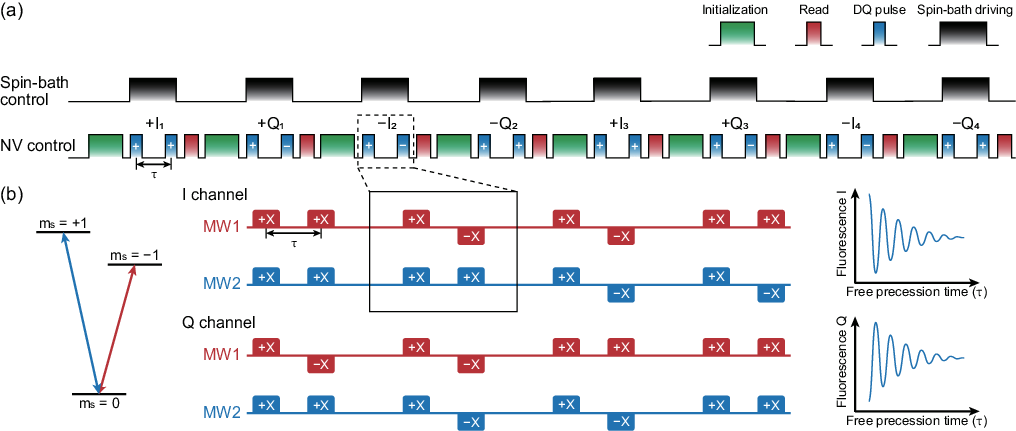}
    \caption{(a) Sequence for magnetic imaging using the DQ 4-Ramsey protocol for NV interrogation together with spin-bath driving. The camera internal exposures alternate between the $I$ and $Q$ channel; on-chip circuitry accumulates differential results between two successive measurements in both channels. For example, the output in the $I$ channel using the sequence in (a) will be $S_{I1}-S_{I2}+S_{I3}-S_{I4}$. (b) Left: Dual-tone MW resonant with $\ket{m_s=0}\leftrightarrow \ket{m_s=+1}$ and $\ket{m_s=0}\leftrightarrow \ket{m_s=-1}$ NV spin transitions is used to implement DQ pulses. Middle: Phase alternations of individual MW tones in a DQ 4-Ramsey sequence. The $\pm\mathrm{X}$ directions are defined in the effective Bloch sphere formed by the two NV spin levels $\ket{m_s=0}$ and $\ket{m_s=+1}$ (or $\ket{m_s=-1}$). The duration of the DQ pulse is the time to change the NV spin state from $\ket{m_s=0}$ to a superposition state with equal population between $\ket{m_s=+1}$ and $\ket{m_s=-1}$. Right: The permutation of MW-phase alternation in the $Q$ channel results in NV fluorescence dynamics of opposite phase to that of the $I$ channel. Differential results between these two measurements are output to an external frame.
    }
    \label{figS2}
\end{figure*}

Figure~\ref{figS2}(a) outlines a particular implementation of the DQ 4-Ramsey protocol in conjunction with spin-bath-driving pulses for magnetic-imaging experiments using the Heliotis heliCam lock-in camera. The NV fluorescence signals from individual DQ Ramsey measurements are stored during the internal exposures of the camera. Additionally, the internal exposures alternate between an in-phase ($I$) and a quadrature ($Q$) channel; and every second measurement in the $I$ and $Q$ channel is subtracted from the first measurement through on-chip circuitry. To obtain the normalized signal $S_{norm}$ in the external frame, we implement a MW-phase alternation scheme as shown in Figure~\ref{figS2}(b).
Two DQ 4-Ramsey sequences are both synchronized with the internal exposures on the $I$ and $Q$ channel. The phase permutation in the $Q$ channel is further chosen to generate a signal $S_{norm}^Q$, with opposite polarity of that in the $I$ channel $S_{norm}^I$. Finally, an external frame containing $S_{norm}$ is obtained from the differences between $S_{norm}^I$ and $S_{norm}^Q$.

For all magnetic imaging experiments (including either with and without spin-bath driving, measurements of spatial magnetic noise, and phantom magnetic-field images), the NV fluorescence signal in the camera external frame $S_{norm}$ is converted to magnetic signal using a magnetometry calibration curve. Briefly, with the Ramsey free evolution interval $\tau$ fixed, the MW frequency is swept around the NV electronic spin resonances to emulate the change of magnetic field $B$; the response of $S_{norm}$ is then measured, producing a magnetometry calibration curve. The optimal MW frequency for magnetic measurements is then determined by maximizing the slope $dS_{norm}/dB$. For DQ magnetometry, the two MWs that are separately resonant with NV electronic spin transitions $\ket{m_s=0}\leftrightarrow \ket{m_s=+1}$ and $\ket{m_s=0}\leftrightarrow \ket{m_s=-1}$ are swept differentially (i.e., a detuning of $\delta$ is applied to one MW tone while a detuning of $-\delta$ is applied to the second MW tone). The optimal choice of the Ramsey free evolution interval $\tau$ is approximately equal to the NV ensemble spin dephasing time $T_2^*$; however, due to Ramsey fringe beating introduced by the hyperfine splitting of the NV spin resonances, $\tau$ is manually adjusted to maximize the median per-pixel magnetic sensitivity $\eta$.

\section{Spin-Bath Driving}
\label{supp:DEER}
Spin-bath driving \cite{bauch2018ultralong} is used to decouple dipolar interactions between the NVs and paramagnetic bath spins ($S=1/2$) in the diamond, including neutral nitrogen $^{15}\mathrm{N}_s^0$ (as we are using $^{15}$N-enriched NV-diamond) and free electrons.  
A strong, multi-tone radio frequency (RF) drive signal is applied throughout the NV layer to induce rapid Rabi oscillations of the bath spins. The effective dipole moments of these spins are then time-averaged to zero, which reduces their inhomogeneous broadening effect on the NV spin resonance
and extends the NV spin dephasing time ($T_2^*$).

\begin{figure}
    \centering
    \includegraphics[width=\linewidth]{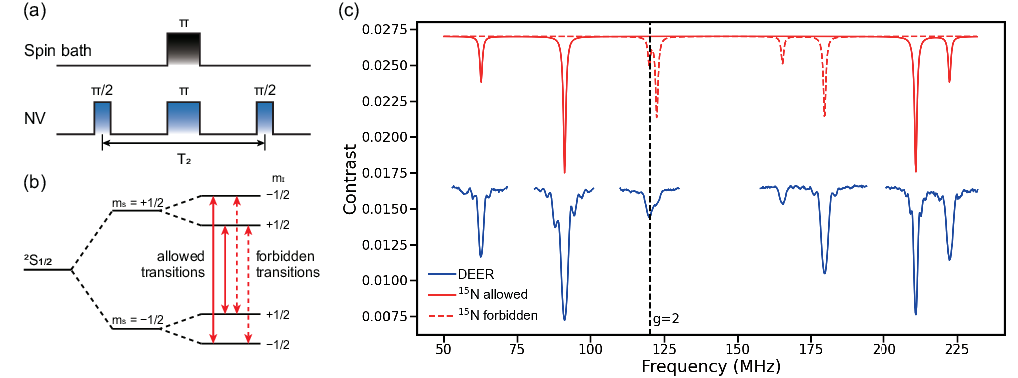}
    \caption{(a) NV double electron-electron resonance (DEER) technique for measuring the $^{15}$N$_s^0$ and free-electron spin resonance spectrum in the QDM diamond sensor. The total free evolution intervals between the initial and final single-quantum (SQ) $\pi/2$ pulse on the NV spins is chosen to match the ensemble NV spin-echo $T_2$ for optimal sensitivity to bath-spin resonances. The frequency of the SQ $\pi$ pulse on the bath spins is swept to measure the NV DEER spectrum by the effect on the NV ensemble fluorescence contrast (see Ref.~\cite{bauch2018ultralong} for details).
    (b) Energy level diagram of the coupled electron-nuclear spin system in $^{15}$N$_s^0$. (c) Simulated $^{15}$N$_s^0$ and free-electron ($\mathrm{g}=2$) spin resonance spectrum (red) overlaid with NV DEER measurement results (blue) and offset vertically for clarity. The simulated resonant frequencies are broadened by a Lorentzian function with linewidth and relative contrast chosen to approximate the measurement data. 
    }
    \label{figS3}
\end{figure}

The frequency components of the RF drive need to be resonant with the various bath-spin transitions, which are experimentally determined using NV double electron-electron resonance (DEER) \cite{de2012controlling, bauch2018ultralong} as shown in Figure~\ref{figS3}(a). To estimate the frequency scanning range for measuring the NV DEER spectrum, we first calculate the expected bath-spin transition frequencies. The free-electron spin transition frequency can be determined from $\gamma_{e}|\vec{B}|$, where $\gamma_{e}$ is the electron gyromagnetic ratio, and $|\vec{B}|$ is the magnitude of the applied bias magnetic field. The transition frequency spectrum for $^{15}\mathrm{N}_s^0$ is calculated using its ground-state Hamiltonian \cite{cox199413c}
\begin{equation}
\label{P1Hamiltonian}
\begin{aligned}
\frac{\hat{H}}{h} =  \frac{\mu_B}{h} \vec{B} \cdot \overline{g} \cdot \vec{\hat{S}} + \frac{\mu_N}{h} \vec{B} \cdot \vec{\hat{I}} + \vec{\hat{S}} \cdot \overline{A} \cdot \vec{\hat{I}},
\end{aligned}
\end{equation}
where $\mu_B$ is Bohr magneton, $h$ is the Planck's constant, $\vec{B} = (B_x,B_y,B_z)$ is the applied bias magnetic field vector, $\overline{g}$ is the electronic g-factor tensor, $\mu_N$ is the Bohr magneton, $\vec{\hat{S}} = (\hat{S}_x,\hat{S}_y,\hat{S}_z)$ is the dimensionless electronic spin vector, $\overline{A}$ is the hyperfine tensor, and $\vec{\hat{I}} = (\hat{I}_x,\hat{I}_y,\hat{I}_z)$ is the dimensionless nuclear spin vector. When the $z$-axis is chosen as one of the four NV crystal axis directions, tensors $\overline{g}$ and $\overline{A}$ are diagonal and defined by
\begin{equation}
\label{P1tensors}
\begin{aligned}
\overline{g} = \begin{bmatrix} g_\perp & 0 & 0\\ 0 & g_\perp & 0 \\ 0 & 0 & g_{||}\end{bmatrix},\,\overline{A} = \begin{bmatrix} A_\perp & 0 & 0\\ 0 & A_\perp & 0 \\ 0 & 0 & A_{||}\end{bmatrix},
\end{aligned}
\end{equation}
where $g_\perp$, $g_{||}$, $A_\perp$, and $A_{||}$ are the gyromagnetic and hyperfine on and off-axis components, respectively. 
$^{15}$N$_s^0$ has $S=1/2$ and $I= 1/2$, leading to the four eigenstates $\ket{m_S=\pm1/2, m_I=\pm1/2}$. The two corresponding dipole-allowed transitions ($\Delta m_S=\pm1$, $\Delta m_I=0$, solid arrows), along with the two first-order forbidden transitions ($\Delta m_S=\pm1$, $\Delta m_I=\pm1$, dashed arrows) are shown in Figure~\ref{figS3}(b). The simulated $^{15}$N$_s^0$ resonance spectrum, using $g_{\perp}=g_{||}=2$, $A_\perp =-559.7$\,MHz, $A_{||}=-113.83$\,MHz, and $|\vec{B}|=4.278$\,mT aligned with an NV axis, is shown in Figure~\ref{figS3}(c), along with the simulated free-electron spin resonance frequency ($\mathrm{g}=2$) and the measured NV DEER spectrum.

After calibration of the RF frequency components, the Rabi driving frequencies associated with the dipole-allowed $^{15}$N$_s^0$ and free-electron transitions are increased to about 2\,MHz, limited by the available power of the amplifier.
The NV dephasing time $T_2^*$ improves with increasing bath-spin Rabi frequency, and is eventually limited by the next dominant dephasing source (NV-NV interactions) \cite{bauch2018ultralong}. The employed RF loop produces spatially-dependent Rabi driving amplitudes, broadening the distribution of $T_2^*$ as indicated in main text Figure~\ref{fig:2}(a).

\section{NV Spin Dephasing Time and Magnetic Sensitivity}
\label{supp:sensorcharacterization}
\begin{figure}
    \centering
    \includegraphics[width=8.5cm]{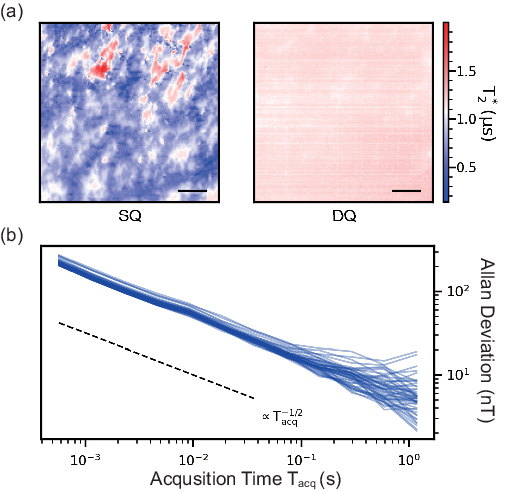}
    \caption{
    (a) QDM maps of NV ensemble dephasing time $T_2^*$ obtained from single-quantum (SQ) and double-quantum (DQ) Ramsey measurements. SQ measurement of $T_2^*$ is limited by heterogeneous strain in the diamond sample. Field of view: $270\times270$\,\textmu m$^2$. Scale bar: 50\,\textmu m.  
    (b) Allan deviations for magnetic noise measurements (no external signal) as a function of acquisition time $T_{acq}$. DQ Ramsey with spin-bath driving protocol is utilized and 50 pixels are randomly chosen within the field of view. A dashed black line depicts power law scaling behavior $\propto T_{acq}^{-1/2}$ as a guide to the eye.}
    \label{figs4}
\end{figure}

The spin-bath driving technique is only effective when the NV spin dephasing time $T_2^*$ is limited by NV interactions with bath spins. The diamond used in the present study has a strain gradient that dominates NV dephasing unless double quantum (DQ) sensing is employed. The strain effect can be seen by comparing $T_2^*$ measured across the NV layer using single-quantum (SQ) and DQ Ramsey sequences; see Figure~\ref{figs4}(a). As discussed in supplementary Section~\ref{supp:DQ4R}, the SQ sensing basis employs a coherence between $\ket{m_s=0}$ and $\ket{m_s=+1}$ (or $\ket{m_s=-1}$), which has a transition frequency dependent on axial crystal stress; while the DQ basis utilizes a coherence between $\ket{m_s=-1}$ and $\ket{m_s=+1}$, with a resonance robust to the axial stress gradient.

Efficient spin-bath driving should result in the per-pixel DQ $T_2^*$ approaching the NV-NV interaction limit due to the high NV density in the employed diamond. Ignoring insignificant per-pixel $T_2^*$ limitations from the 0.005\% $^{13}$C, $T_2^*\{^{13}\mathrm{C}\}=200$\,\textmu s, and the $\sim$\,0.013\,\textmu T/\textmu m bias magnetic field gradient, $T_2^*\{\mathrm{bias}\}=850$\,\textmu s, we estimate the NV-NV interaction limit as \cite{barry2020report}

\begin{equation}
\begin{aligned}
\frac{1}{T_2^*} \approx &\frac{1}{T_2^*\{\mathrm{NV}\}_{||}}+\frac{1}{T_2^*\{\mathrm{NV}\}_{\nparallel}}\\
\approx&A_{\mathrm{NV}_{||}}[\mathrm{NV}_{||}]+\zeta_{\nparallel}A_{\mathrm{NV}_{\nparallel}}[\mathrm{NV}_{\nparallel}].
\end{aligned}
\end{equation}
Here $[\mathrm{NV}]_{||}$ is the concentration of NV spins in the same crystal axis group used for sensing and $[\mathrm{NV}]_{\nparallel}$ is the concentration of NV spins in other groups. Since the bias magnetic field is aligned to one of the four NV crystal axes, $[\mathrm{NV}]_{||}=\frac{1}{3}[\mathrm{NV}]_{\nparallel}=0.59$\,ppm \cite{edmonds2021characterisation}. The constants $A_{\mathrm{NV}_{||}}=0.247$\,\textmu s$^{-1}$$\cdot$ppm$^{-1}$ and $A_{\mathrm{NV}_{\nparallel}}=0.165$\,\textmu s$^{-1}$$\cdot$ppm$^{-1}$ \cite{barry2020report} characterize the dipolar interaction strength for pairs of NV spins in the same and different groups, respectively. $\zeta_{\nparallel}$ is a dimensionless factor of order unity accounting for imperfect initialization of NV$_{\nparallel}$ into the $\ket{m_s=0}$ state. Experimental determination of $\zeta_{\nparallel}$ is complicated by unequal laser polarization projected among the three NV$_{\nparallel}$ axes in a total internal reflection excitation scheme (supplementary Section~\ref{supp:exp}). As an estimation for the NV-NV interaction limited $T_2^*$, we provide the results for perfect ($\zeta_{\nparallel}=0$) and partial ($\zeta_{\nparallel}=0.5$) initialization cases, thus $T_2^*(\zeta_{\nparallel}=0)=6.6$\,\textmu s and $T_2^*(\zeta_{\nparallel}=0.5)=3.4$\,\textmu s. In addition, accounting for the $2\times$ sensitivity to the magnetic noise in a DQ scheme ($T_2^*(\mathrm{DQ})=\frac{1}{2}T_2^*$) \cite{bauch2018ultralong}, we finally expect the NV-NV interaction limited $T_2^*$ using DQ measurement to be approximately 2.5\,\textmu s, the average of $T_2^*(\zeta_{\nparallel}=0,\mathrm{DQ})$ and $T_2^*(\zeta_{\nparallel}=0.5,\mathrm{DQ})$, which is reflected in the results shown in main text Figure~\ref{fig:2}(b). Experimental per-pixel $T_2^*$ with and without spin-bath driving are extracted by fitting the measured Ramsey fringes to a sum of oscillations with a common Lorentzian exponential decay \cite{Hart2021-4Ramsey}. The uncertainties reported in main text for the median per-pixel $T_2^*$ are estimated from fitting uncertainties using the covariance matrices returned by the PYTHON \texttt{scipy.optimize.curve\_fit} function.

The experimental per-pixel magnetic sensitivity results are computed from $\eta=\delta B\sqrt{T_{acq}}$, where $\delta B$ is the smallest magnetic field that can be measured with $\mathrm{SNR}=1$ after $T_{acq}$ acquisition time. For a camera operating at an external frame rate $F_s$, a series of magnetic noise images are collected. Then the magnetic noise of each pixel, $\sigma_{pxl}$, is obtained from the standard deviation of the magnetometry data from that pixel across all recorded frames. The per-pixel magnetic sensitivity can then be written as
\begin{equation}
\label{eq:sensitivitypractical}
\begin{aligned}
\eta =  \frac{\sigma_{pxl}}{\sqrt{F_s}}.
\end{aligned}
\end{equation}
The results displayed in main text Figure~\ref{fig:2}(a) are obtained using 500 frames collected for $F_s\approx1.4$\,kHz with spin-bath driving and $F_s\approx1.7$\,kHz without spin-bath driving. The uncertainty reported for the measured median per-pixel sensitivity $\eta$ with spin-bath driving is computed as the standard deviation of $\eta$ across 15 repeated experiments, each consisting of 500 frames. The experiment without spin-bath driving is only repeated once. As an estimate of the uncertainty for $\eta$ without spin-bath driving, the 500 frames are divided into ten 50-frame sets and the standard deviation of $\eta$ across the ten frame sets is reported. Experiments for per-pixel magnetic sensitivity measurements with spin-bath driving are also performed at different values of $F_s$ ($\approx$\,3.1\,kHz, 2.2\,kHz, 1.0\,kHz and 0.5\,kHz) to investigate potential technical noise at different imaging speeds. Across all these measurements, the difference between the maximal and minimal median magnetic sensitivity values is $\lesssim$\,5\,\%. The Allan deviation for DQ Ramsey magnetic noise measurements (i.e., no external signal) with $F_s\approx1.4$\,kHz and spin-bath driving is shown in Figure~\ref{figs4}(b), using 7500 frames. Note that the overhead time associated with data transfer of the 15 batches of 500 frames (approximately 5\,min, see supplementary Section~\ref{supp:exp}) is neglected in $T_{acq}$ when computing the Allan deviation, similar to Refs.~\onlinecite{Hart2021-4Ramsey, Marshall2022precision}. 

The photon shot-noise-limited magnetic sensitivity for NV Ramsey magnetometry (either SQ or DQ) is given by \cite{barry2020report} 

\begin{equation}
\label{suppeq:RamseyPhotonShotSensitivity}
\begin{aligned}
\eta_{shot} = \frac{\hbar}{\Delta m_s g_e \mu_B} \frac{1}{Ce^{(-\tau/T_2^*)^p}\sqrt{N}}\frac{\sqrt{t_{I,R}+\tau}}{\tau},
\end{aligned}
\end{equation}
where $g_e\approx2.003$ is the NV center’s electronic $g$ factor, $\mu_B$ is the Bohr magneton, $\hbar$ is the reduced Planck constant, $\Delta m$ accounts for the difference of $m_s$ states used for sensing, $C$ is the optical measurement contrast at Ramsey free evolution time $\tau=0$, $T_2^*$ is the NV spin dephasing time, $p$ is a parameter used to describe the Ramsey envelope decay shape, $N$ is the average number of photons collected per measurement, and $t_{I,R}$ is the combined initialization and readout time during the Ramsey sequence. We use the following parameters to give an estimate of $\eta_{shot}$ for a single QDM pixel using the present NV-diamond sensor. In a DQ sensing scheme, $\Delta m_s=2$. By focusing the total NV fluorescence within the field of view onto a photodiode, we experimentally determine the ensemble-averaged Ramsey oscillation parameters to be about $C = 3$\,\%, $T_2^*= 2.3$\,\textmu s, and $p=1$. We use $t_{I,R} =7.04$\,\textmu s and $\tau=2.29$\,\textmu s, typical for a QDM magnetic imaging experiment, and estimate $N=4\times 10^4$ photons are collected in a single camera pixel (detailed in the next paragraph) during the $2.15$\,\textmu s camera readout window. Using these values, the estimated photon shot-noise-limited per-pixel magnetic sensitivity is 
$1.7$\,nT$/\sqrt{\mathrm{Hz}}$.
\begin{figure*}
    \centering
    \includegraphics[width=8.5cm]{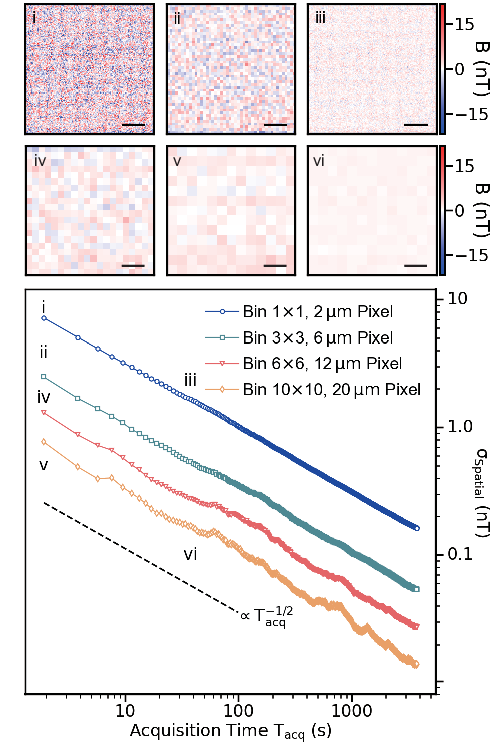}
    \caption{Additional pixel binning results of spatial magnetic noise floor in QDM images. Measurement data are obtained using the differential measurement protocol. Upper six panels: Images of the background magnetic noise with different acquisition times and pixel binning (values indicated in lower panel). Scale bar: 50\,\textmu m. Lower panel: Spatial magnetic noise floor $\sigma_{spatial}$ as a function of acquisition time $T_{acq}$. A dashed black line depicts power law scaling behavior $\propto T_{acq}^{-1/2}$ as a guide to the eye.}
    \label{figS5}
\end{figure*}
As discussed in supplementary Section~\ref{supp:DQ4R}, the camera accumulates a differential photon number result $\Delta N$ between two successive internal exposures. When outputting an external frame, $\Delta N$ is additionally converted to the camera device unit (DU) via a conversion factor (CF), giving $\mathrm{DU}=\mathrm{CF}\Delta N$.
To calibrate CF and infer the photon number $N$ collected by a single pixel during the $2.15$\,\textmu s camera exposure window, 
we perform two sets of measurements to extract the photon noise and signal scaling behavior as a function of the camera readout duration $t$.
For the noise measurement, we vary $t$ while ensuring the readout times for two successive camera internal exposures are identical, thus the mean $\Delta N=0$. However, the characteristic standard deviation (std) in the device unit for each pixel scales as 
\begin{equation}
\label{suppeq:noisescale}
(\mathrm{std})^2=A_{noise}t=(\mathrm{CF})^22N,
\end{equation}
where $A_{noise}$ is the per-pixel noise scaling slope factor to be determined. For the signal measurement, we vary the exposure time difference $\Delta t$ between two successive camera internal exposures, thus the signal $S$ in terms of the camera  device unit is
\begin{equation}
\label{suppeq:signalscale}
S=A_{signal}\Delta t=\mathrm{CF}\Delta N,
\end{equation}
where $A_{signal}$ is the per-pixel signal scaling slope factor to be determined. Further recognizing that $\Delta N/N=\Delta t/t$, the CF can be deduced by
\begin{equation}
\label{suppeq:signalscale}
\mathrm{CF}=\frac{A_{noise}}{2A_{signal}}.
\end{equation}
For the camera operating at unity gain and an internal exposure rate matching the Ramsey sequence readout intervals, we use the measured photon noise and signal scaling behavior to find an average $\mathrm{CF}\approx\frac{1}{446}$ across the field of view.
By further recording the DU as a function of $\Delta t$, an estimated photon number of $N=4\times 10^4$ is collected in a single pixel after a 2.15\,\textmu s readout duration, for typical QDM operating parameters (detailed above). We note that $N$ is technically the photoelectron number, which differs by a multiplicative factor from the actual photon number collected by a camera pixel (about 70\,\% optical fill factor and 90\,\% quantum efficiency around the NV fluorescence wavelengths for the Heliotis camera with micro-lenses).

The quantization noise of the Heliotis camera includes electronic and digitization noise. For each external frame, the noise is dependent on the cycles ($N_c$) spent accumulating differences between successive internal exposures, and can be estimated in DU as \cite{heliotis}
\begin{equation}
\label{suppeq:quantizationnoise}
\sqrt{0.81+0.16N_c}.
\end{equation}  
Converting the estimate in DU to equivalent magnetic noise and including the contributions from both the $I$ and $Q$ readout channels (supplementary Section~\ref{supp:DQ4R}), the estimated quantization-noise-limited per-pixel magnetic sensitivity is $2.7$\,nT$/\sqrt{\mathrm{Hz}}$. After adding photon shot-noise and camera quantization noise in quadrature, the combined estimate of per-pixel magnetic sensitivity is $3.2$\,nT$/\sqrt{\mathrm{Hz}}$. As discussed in main text, the measured median per-pixel magnetic sensitivity is 4.1(1)\,nT$/\sqrt{\mathrm{Hz}}$.

In future work, the magnetic sensitivity limit due to quantization noise 
can be reduced for a QDM using the Heliotis camera by increasing the NV fluorescence intensity, NV spin-state readout contrast, and NV spin dephasing time. The dynamic range of the Heliotis camera is fixed (DU from 0 to 1023, 10-bit), and the camera sensor is optimized for unity gain \cite{heliotis}. For a given accumulation cycle $N_c$ of internal exposures, the quantization noise is a constant in DU (see Equation~\ref{suppeq:quantizationnoise}). Thus, if the maximally achievable slope of the magnetometry curve (supplementary Section~\ref{supp:DQ4R}), which is used for converting the camera external frame output from DU to a magnetic-field value, can be increased, then the magnetic-noise contribution from quantization noise will decrease. The maximized slope of a magnetometry curve for a given $N_c$ is proportional to both NV fluorescence intensity and spin-state readout contrast (which affect the accumulated differential photon number $\Delta N$ in a camera external frame, see above); and also to NV spin dephasing time $T_2^*$ (which affects the optimal sensing interval).
Hence, further diamond engineering --- e.g., permitting the use of increased laser illumination intensity while mitigating the contrast degradation due to NV charge-state conversion \cite{edmonds2021characterisation}, and also improving the NV spin dephasing time --- is beneficial for mitigating the quantization-noise limit on magnetic sensitivity.

\begin{figure}
    \centering
    \includegraphics[width=8.5cm]{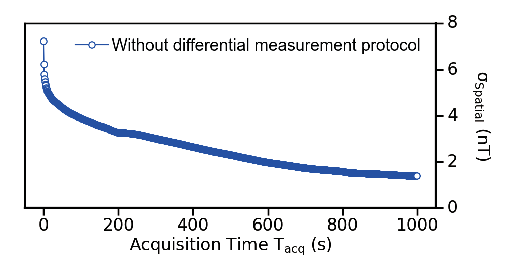}
    \caption{Spatial magnetic noise floor $\sigma_{spatial}$ as a function of acquisition time $T_{acq}$. The differential measurement protocol is not employed and thus the utility of time averaging is limited. No pixel binning is applied.}
    \label{figS6}
\end{figure}

\section{Differential Measurement Protocol}
\label{supp:diffmeas}

As described in main text, we use a differential measurement protocol to mitigate the spatial-temporal variation of magnetic noise. This protocol takes the difference of measurement pairs, with each pair consisting of two sets of 500 frames acquired with identical magnetic sensing sequences --- except for the currents applied to the phantom being modulated. 
For static magnetic-field measurements, one set of frames is collected with positive current applied to the phantom, while the other frame set is collected with negative current. For dynamic magnetic-field measurements, the applied current is modulated between on and off. 
The first modulation scheme has identical magnetic sensitivity compared to measuring with steady (unmodulated) current flowing through the phantom, while the second modulation scheme costs approximately a factor of 2 in sensitivity.
The camera external frame rates for imaging both static and dynamic magnetic fields (as well as for measuring the scaling of the time-averaged spatial magnetic noise floor in main text Figure~\ref{fig:2}(b)) are set to about 528\,Hz (except for the sub-millisecond magnetic-imaging experiment described in supplementary Section~\ref{supp:subms}); thus the acquisition times for both sets of 500 frames in the measurement pair are $\sim$\,1\,s. However, considering the overhead associated with data transfer (supplementary Section~\ref{supp:exp}), the applied current is modulated on a timescale $\sim$\,10\,s.

After obtaining the differential results of each measurement pair, the magnetic data is further processed by subtracting the image resulting from a 2D Gaussian spatial filter (kernel size $\sigma_{x,y}^{filt}=18$\,\textmu m) in order to reduce the effect of any residual, large-length-scale magnetic gradients across the NV layer. 
The kernel size is experimentally determined to be larger than the variation length scale of the phantom magnetic field. Finally, the processed magnetic measurement data are averaged over the acquisition time.

Characterization of the spatial magnetic noise floor in QDM images are reported in Figure~\ref{figS5} for alternative pixel binning, similar to the procedure and presentation in main text Figure~\ref{fig:2}(b).

\section{Spatial Magnetic Noise Floor without Differential Measurement Protocol}
\label{supp:notmodulated}
To determine the spatial magnetic noise floor without using the differential measurement protocol, we average all of the successively acquired magnetic frame sets for a given acquisition time $T_{acq}$. Before averaging, each acquired frame set is spatially filtered with a Gaussian kernel as described in supplementary Section~\ref{supp:diffmeas} to reduce the effect of background magnetic gradients, and subtracted from the static pixel-to-pixel baseline difference pattern (which arises from manufacturing variations in the camera). The results are shown in Figure~\ref{figS6}.

\begin{figure}
    \centering
    \includegraphics[width=8.5cm]{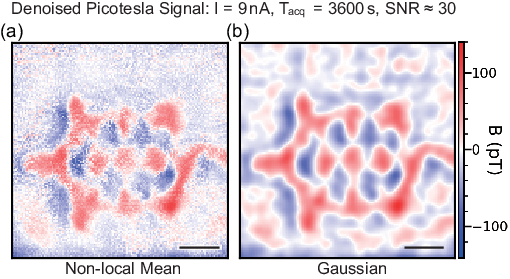}
    \caption{Denoised QDM picotesla-scale magnetic-field images after about 3600\,s signal acquisition time. Original measurement data is presented in main text Figure~\ref{fig:3}(b), left panel. Here, two denoising methods, Non-local mean (NLM) and Gaussian smoothing, are applied. Both algorithms are successful at improving $\mathrm{SNR}$. However, NLM denoising, as shown in (a), additionally excels at restoring fine structures (e.g., contours, patterns, and edges where the Laplacian is non-zero) from a noisy image that can be otherwise distorted by a Gaussian smoothing, as shown in (b). Scale bar: 50\,\textmu m}
    \label{figS7}
\end{figure}

\section{Denoising of Picotesla-Scale Magnetic Image}
\label{supp:denoise}

A non-local mean (NLM) denoising algorithm \cite{buades2011non} is applied to the picotesla-scale magnetic image shown in main text Figure~\ref{fig:3}(b), right panel, to improve $\mathrm{SNR}$. The main feature of this denoising algorithm is the ability to restore both the large-scale geometries and fine structures (e.g., contours, patterns, and edges where the Laplacian is non-zero) from a noisy image; whereas local-mean filtering, such as Gaussian smoothing, can blur fine structures \cite{buades2005non}. Briefly, to denoise a pixel $p$, the NLM algorithm will compute the weighted averages of all other pixels $q$ within a search window centered at pixel $p$. The weight assigned to a pixel $q$ is computed using the Euclidean $L^2$ distance between two template windows centered at $q$ and $p$, respectively --- i.e., summation over squared pixel-pair value differences with each pair consisting of two pixels from the template windows around $q$ and $p$, respectively --- and normalized using a 1D Gaussian function. If the template window around $q$ has a similar intensity profile compared to that of $p$, a higher weight will be assigned to $q$; this process is then repeated for all $q$ within the search window centered at $p$. Thus the denoised pixel $p$ is predominantly the averaged value of similar $q$ pixels, rather than its immediate neighbors in a local-mean denoising method.

For QDM magnetic imaging of the fabricated wire phantom, regions of similar magnetic-field patterns are not necessarily adjacent to each other. We expect that NLM denoising can increase $\mathrm{SNR}$ while reducing distortions of the magnetic-field pattern, as compared to local-mean methods such as a Gaussian smoothing. The sizes of the search and template windows in NLM denoising are set to $21\times21$\,pixels and $7\times7$\,pixels, respectively; taken as the recommended the values from Ref.~\cite{buades2011non}. The kernel size for the 1D Gaussian normalization function is about 44\,pT, $\sim$\,0.5\,$\times$ of the spatial magnetic noise floor $\sigma_{spatial}$ in the picotesla-scale magnetic image after about 3600\,s signal acquisition time ($T_{acq}$), as shown in the left panel of main text Figure~\ref{fig:3}(b). We manually optimize this kernel size to improve $\mathrm{SNR}$ to $\sim$\,30. The NLM denoised results are shown in main text Figure~\ref{fig:3}(b), right panel, and reproduced in Figure~\ref{figS7}(a). For Gaussian smoothing in comparison, the measured picotesla-scale magnetic image is convolved with a Gaussian filter and its kernel size $h$ is set to about 2.6\,pixels to increase SNR of the denoised image to $\sim$\,30. For pure additive Gaussian white noise in an image, the standard deviation of the noise reduces as $\frac{1}{h\sqrt{8\pi}}$ \cite{buades2005review}; and we thus expect $h\approx2.9$\,pixel to reach the desired $\mathrm{SNR}$ --- similar to the results found for NLM denoising.
As shown in Figure~\ref{figS7}(b), the contours of Gaussian-smoothed magnetic-field patterns are qualitatively more distorted than the NLM results. A quantitative perspective to evaluate the performance of the denoising algorithms is comparing the mean square errors (MSE) \cite{buades2011non}, obtained using the Euclidean $L^2$ distance between the original and the denoised images. The NLM denoising method produces about 17\,\% smaller MSE than the Gaussian smoothing result.

\begin{figure}
    \centering
    \includegraphics[width=\linewidth]{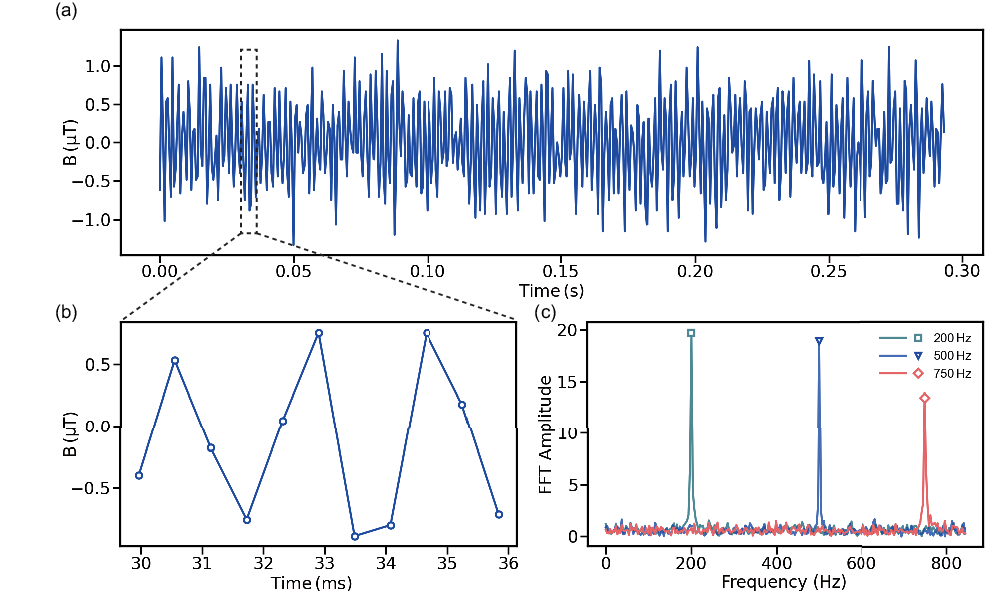}
    \caption{QDM magnetic imaging with sub-millisecond time resolution. Camera external frame rate is set to $F_s\approx1.7$\,kHz  (i.e., time step between frames of about 0.6\,ms). Magnetic signals are produced by applying single-tone AC signal voltages to the phantom.  AC frequencies of 200\,Hz, 500\,Hz, and 750\,Hz are used for different runs of the demonstration experiment. (a, b) Example time-domain magnetic-measurement data (500\,Hz signal) from a single pixel in the QDM image, selected to have significant amplitude variation over a signal cycle. (c) Time-domain data from the single pixel is Fourier transformed and displayed, here for all three AC frequencies used in the demonstration experiments. The resulting spectrum is in good agreement with the applied signal frequencies; and with a simulation of the QDM measurement protocol that leads to a reduction in spectral peak amplitude near the Nyquist frequency ($F_s/2$).}
    \label{figS8}
\end{figure}

\section{Magnetic Imaging with Sub-Millisecond Time Resolution}
\label{supp:subms}
To demonstrate the capability of the present QDM for magnetic imaging with sub-millisecond time resolution, we operate the Heliotis camera at an external frame rate of $F_s\approx1.7$\,kHz. A single-tone AC signal voltage of fixed amplitude is then applied to the phantom; and 500 dynamic magnetic images are acquired (over $\sim$\,290\,ms), with a time step between images $=1/F_s\approx0.6$\,ms. The experiment is conducted for three different runs with signal frequencies of 200\,Hz, 500\,Hz, and 750\,Hz; and each run is repeated once. A single pixel (no binning), selected to have significant amplitude variation over a signal cycle, is used to characterize the time-domain oscillations of the magnetic-field measurements (see Figure~\ref{figS8}(a, b)); the time-domain data is then
Fourier-transformed
to identify the signal frequency components, as shown in Figure~\ref{figS8}(c).
The results are in good agreement with the signal frequencies. However, suppression of the spectral peak amplitude is observed near the Nyquist frequency ($F_s/2$), due to the effective low-pass filtering in generating a camera external frame, as each external frame contains the accumulated photoelectrons from multiple internal exposures for magnetic-signal measurements using the DQ 4-Ramsey sequence (supplementary Section~\ref{supp:DQ4R}). In all the magnetic imaging data acquired in this sub-millisecond demonstration, the internal exposure (i.e., readout of NV fluorescence after a DQ Ramsey sequence) rate is 122\,kHz; and a single external frame is produced by the camera after digitizing the cumulative photoeletrons from 72 internal exposures, giving a time step between external frames of about 0.6\,ms irrespective of the AC signal frequency. The accumulation cycles of internal exposures can be reduced to increase the temporal resolution of external frames, ultimately limited by the Heliotis camera's maximal external frame rate ($\sim$\,3.8\,kHz). A simulation of this low-pass filtering effect in camera external frames — using synthesized data sets for each of the AC signal frequencies, sampling at the internal exposure rate, averaging over the accumulation cycles for the time-domain data, and then a Fourier transform — is in good agreement with the experimental measurements, as shown in Figure~\ref{figS8}(c).

\label{supp:projection}
\begin{figure}
    \centering
    \includegraphics[width=8.5cm]{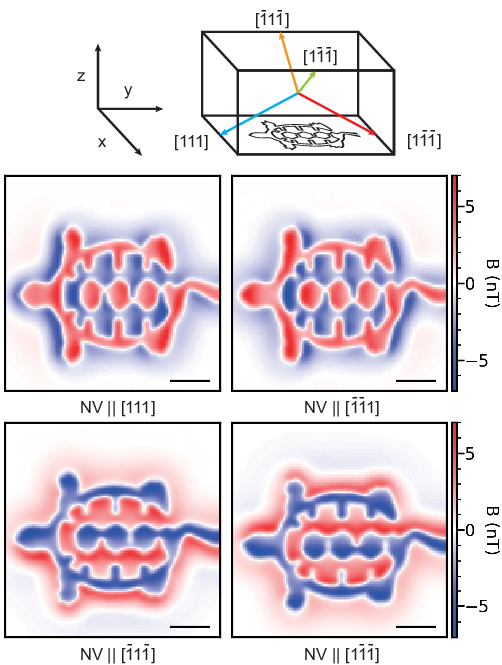}
    \caption{Simulated phantom static magnetic-field patterns (Bottom four panels) projected along four NV axes in the diamond crystal (shown schematically in the top panel). In the present study, the QDM is configured to image the magnetic field projected along the $\mathrm{[111]}$ axis. The simulation procedure is detailed in the caption for Figure~\ref{fig:3}(a), left panel, in main text.}
    \label{figS9}
\end{figure}

\section{Phantom Magnetic-Field Patterns Projected along Different NV Axes}
\label{supp:fouraxes}
The diamond employed in the present QDM is cut into a square prism with side facets perpendicular to the $\mathrm{[110]}$ and $[1\overline{1}0]$ crystal axes. This geometry facilities alignment of the diamond sensor to the fabricated wire phantom; the diamond is then directly mounted onto the silicon carbide wafer that supports the fabricated phantom structure to fix their relative orientation. All magnetic imaging in the present study is of the projection of the phantom magnetic-field pattern along one of the four NV quantization axes in the diamond, corresponding to the $\mathrm{[111]}$ crystal axis. Figure~\ref{figS9} shows the simulated phantom static magnetic-field patterns projected along all four NV crystal axes.

\end{document}